\documentclass[fleqn,usenatbib]{mnras}

\usepackage{newtxtext,newtxmath}

\usepackage[T1]{fontenc}
\usepackage{ae,aecompl}

\usepackage{multirow,multicol}
\usepackage[dvips]{graphicx}   
\usepackage[dvipsnames]{xcolor}
\usepackage[normalem]{ulem}
\usepackage{hyperref}
\usepackage{breakurl}

\DeclareRobustCommand{\ion}[2]{%
\relax\ifmmode
 \ifx\testbx\f@series
  {\mathbf{#1\,\mathsc{#2}}}\else
  {\mathrm{#1\,\mathsc{#2}}}\fi
 \else\textup{#1\,{\mdseries\textsc{#2}}}%
 \fi}

\newcommand{\logNHIcm}{\ensuremath{\log({\rm N_{H\, \textsc{i}}\: /\: cm}^{-2})}}
\newcommand{\MsunYr}{\ensuremath{\textrm{M$_{\odot}$~yr$^{-1}$}}}

\newcommand{\HST}{\emph{HST}}
\newcommand{\lya}{\mbox{${\rm Ly}\alpha$}}
\newcommand{\HI}{\ion{H}{i}}

\title[Absorption-selected galaxies at $z\sim 2.3$]{Absorption-selected galaxies trace the low-mass, late-type, star-forming population at $z\sim 2-3$}

\author[Rhodin, N. H. P. et al.]
       {N. H. P. Rhodin\thanks{nhp.rhodin@gmail.com}$^1$,
         J.-K. Krogager$^2$,
         L. Christensen$^1$,
         F. Valentino$^3$,
         K.~E. Heintz$^{3,\,4}$,
         \newauthor 
         P. M\o ller$^{5,\,6}$,
         T. Zafar$^7$,
         J.~P.~U. Fynbo$^3$,\\         
    $^1$ DARK, Niels Bohr Institute, University of Copenhagen, Jagtvej 128, DK-2200 Copenhagen N, Denmark\\
    $^2$ Institut d'Astrophysique de Paris, CNRS-UPMC, UMR7095, 98bis bd Arago, 75014 Paris, France\\
    $^3$ Cosmic Dawn Center (DAWN), Niels Bohr Institute, University of Copenhagen, Jagtvej 128, DK-2200 Copenhagen N, Denmark\\
    $^4$ Centre for Astrophysics and Cosmology, Science Institute, University of Iceland, Dunhagi 5, 107 Reykjav\'ik, Iceland\\
    $^5$ European Southern Observatory, Karl-Schwarzchildstrasse 2, 85748 Garching bei M\"unchen, Germany\\
    $^6$ Niels Bohr Institute, University of Copenhagen, Jagtvej 128, DK-2200 Copenhagen N, Denmark\\
    $^7$ Australian Astronomical Optics, Macquarie University, 105 Delhi Road, North Ryde, NSW 2113, Australia
       }
       
\date{Received ; accepted }

\begin{document}

\maketitle

\label{firstpage}

\begin{abstract}
We report on the stellar content, half-light radii and star formation
rates of a sample of 10 known high-redshift ($z\gtrsim 2$) galaxies
selected on strong neutral hydrogen (\ion{H}{i}) absorption
($\logNHIcm > 19$) toward background quasars.  We use observations
from the {\it Hubble Space Telescope} (\HST) Wide Field Camera 3 in
three broad-band filters to study the spectral energy distribution
(SED) of the galaxies. Using careful quasar point spread function
subtraction, we study their galactic environments, and perform the
first systematic morphological characterisation of such
absorption-selected galaxies at high redshifts. Our analysis reveals
complex, irregular hosts with multiple star-forming clumps. At a
spatial sampling of 0.067~arcsec per pixel (corresponding to 0.55~kpc
at the median redshift of our sample), 40\% of our sample requires
multiple S\'ersic components for an accurate modelling of the observed
light distributions.  Placed on the mass--size relation and the `main
sequence' of star-forming galaxies, we find that absorption-selected
galaxies at high redshift extend known relations determined from deep
luminosity-selected surveys to an order of magnitude lower stellar
mass, with objects primarily composed of star-forming, late-type
galaxies.  We measure half-light radii in the range $r_{1/2} \sim 0.4$
to $2.6$~kpc based on the reddest band (F160W) to trace the oldest
stellar populations, and stellar masses in the range $\log
(\mathrm{M}_{\star}/\mathrm{M}_{\odot}) \sim 8$ to $10$ derived from
fits to the broad-band SED. Spectroscopic and SED-based star formation
rates are broadly consistent, and lie in the range $\log
(\mathrm{SFR}/\MsunYr) \sim 0.0$ to $1.7$.
\end{abstract}


\section{Introduction}
\label{sec:intro}

Galaxies can be selected via their gas cross-section when there is a
chance alignment with a background quasar along the line of sight.  In
neutral hydrogen (\ion{H}{i}), the most \ion{H}{i}-rich absorbers are
the Damped Lyman-$\alpha$ Absorbers (DLAs; $\logNHIcm \geq 20.3$;
\citealt{Wolfe1986}) and sub-DLAs ($19.0 \leq \logNHIcm < 20.3$; e.g.,
\citealt{Peroux2003, Zafar2013}).  Such high column densities imprint
deep \HI\ absorption lines with characteristic Lorentzian damping
wings on the quasar spectrum, and are always accompanied by
low-ionisation metal line complexes \citep{Prochaska2003,
  Noterdaeme2012, Rafelski2014}.  Unless otherwise specified, sub-DLAs
and DLAs will collectively be referred to as \emph{strong}
\HI\ \emph{absorbers} throughout this work.

DLAs alone account for $>80\,$\% of the neutral gas out to redshift
$z\sim 5$ \citep{Prochaska2005, Noterdaeme2012, Crighton2015,
  SanchezRamirez2016}, and combined with the contribution from
sub-DLAs, strong \ion{H}{i} absorbers effectively probe the neutral,
chemically enriched gaseous environments of galaxies. The connection
that strong \ion{H}{i} absorbers hold to their harbouring galaxies can
be studied by correlating absorption properties with complementary
information of the host in emission. Such analyses suggest that
absorption-selected galaxies are consistent with the faint end of
Lyman-break galaxies \citep{Moller2002}, and that they probe a more
representative portion (low mass, faint end of the luminosity
function) of galaxy populations across cosmic time than conventional
luminosity selections \citep{Moller2002, Fynbo2008, Berry2016,
  Krogager2017}.

A recent string of observations indicates that galaxies associated
with strong \HI\ absorbers at $z\sim 0.7$ exhibit suppressed star
formation rates (SFRs) compared to the stellar-mass--SFR relation at
this redshift \citep{Kanekar2018, Moller2018, Rhodin2018}.  Such a
suppression of the SFR at a given stellar mass is surprising since
\HI\ absorption-selected galaxies are thought to trace gas-rich and
actively star-forming galaxies \citep{Moller2002, Fynbo2008,
  Krogager2017}.  The difference observed at low-redshift may be due
to redshift-evolution in the cross-section of \HI\ gas leading to
different samples of absorption-selected galaxies
\citep[e.g.][]{Rhodin2019}; However, samples of high-redshift galaxies
with stellar mass measurements are currently too small to draw
meaningful conclusions.

At high redshifts ($z\gtrsim 2$), the small sample size is caused by a
combination of low angular separation and a high brightness contrast
between target galaxy and background quasar; an increased surface
brightness dimming with redshift; and the lower mean mass (and
therefore luminosity) of galaxies selected by gas cross-section.
Indeed, the lack of detections and reported survey statistics are
consistent with scaling relation arguments, which suggest that the
emission-line targets often fall below detection-limits in blind
surveys; whilst pre-selecting on the absorption metallicity yields
higher detection rates \citep{Fynbo2010,Fynbo2011} as metal-rich
galaxies tend to be more massive, and therefore more luminous
\citep{Krogager2017}.

Observations at $z\sim 2-3$ allow for simultaneous measurements of the
neutral hydrogen column density from the damped \lya\ absorption
profile and detailed absorption analysis of metal lines with
ground-based spectroscopy \citep{Noterdaeme2012}. This has ensured
high fidelity data to base followup campaigns on in search of the
counterparts in emission. Indeed, the damped \lya\ absorption trough
effectively blocks out the quasar light and can be used to search for
\lya\ emission from the hypothesised host. Owing to the resonance
nature of the hydrogen \lya\ line, which is known to affect the
emerging line-flux from high-$z$ galaxy populations
\citep{Verhamme2008,Laursen2009,Hayes2010} and to its efficient
destruction by dust in more chemically enriched, massive and luminous
galaxies (which could mitigate any selection on absorption
metallicity), such Ly$\alpha$ searches often resulted in
non-detections of absorber counterparts \citep{Fynbo2011, Fynbo2013,
  Krogager2017}.  However, taking advantage of the large wavelength
coverage in modern spectrographs to simultaneously detect
strong-rest-frame optical emission lines, ground-based observational
efforts have become increasingly successful, most prominently seen in
the high detection-rate achieved with the Very Large Telescope (VLT)
X-Shooter \citep[VLT/X-Shooter,][]{Vernet2011} campaign
\citep{Fynbo2010,Krogager2017}.

Whereas spectroscopic searches for galaxy counterparts in the past
primarily focused on identifying the correct host, analyses of the
emission originating from their stellar components were typically
reported as individual case studies
\citep{Moller2002,Fynbo2013,Krogager2013,Augustin2018}. Even though we
have assembled a significant sample of $z\gtrsim2$ absorber--galaxy
pairs, a comprehensive investigation of their stellar properties is
lacking.  Here, we take advantage of the detailed information
available from ground-based spectra to search for the counterparts of
strong \ion{H}{i} absorbers at $z \sim 2-3$ using multi-band imaging
from the \emph{Hubble Space Telescope} (\emph{HST}).  With its
exquisite spatial resolution and in absence of the Earth's atmosphere,
\emph{HST} allows us to disentangle the emission from the intervening
galaxy and background quasar and, thereby, determine stellar masses
and sizes. This directly addresses the low-number statistics; enables
the exploration of scaling relations at high-$z$ independently of
samples at low redshift; and allows us to probe the low-mass
extensions of any relations established from luminosity-selected
galaxy samples.

The paper is organised as follows: Sect. \ref{sec:obs} describes our
sample selection, observations, and data-reduction;
Sect. \ref{sec:results} presents our imaging, photometry,
morphological analysis, and spectral energy distribution (SED) based
stellar mass measurements; and Section \ref{sec:discussion} places our
findings in context with known high-redshift galaxy scaling relations.
In Sect.~\ref{sec:conclusions}, we summarise our conclusions.
Throughout this paper, we assume a flat $\Lambda$ cold dark matter
($\Lambda$CDM) cosmology, with $\mathrm{H}_0 =
70.4~\mathrm{km~s}^{-1}~\mathrm{Mpc}^{-1}$ and $\Omega_{\Lambda} =
0.727$ \citep{Komatsu2011} to ensure consistency with prior work
\citep[e.g.][]{Christensen2014,Moller2018,Rhodin2018}.  Magnitudes are
reported in the AB magnitude system. Star formation rates and stellar
masses are derived using the \cite{Chabrier2003} initial mass function
(IMF). To quantify the degree of consistency between two measurements,
$\mu _1 \pm \sigma _1$ and $\mu _2 \pm \sigma _2$, we report the
number of sigmas statistical tension as $\lvert \mu _1 - \mu _2 \rvert
/ \sqrt{\sigma _1 ^2 + \sigma _2 ^2}$. All logarithmic quantities are
expressed in base 10.

\renewcommand{\arraystretch}{1.50}

\begin{table*}
\caption{Log of \emph{HST} observations.}
\label{tab:obslog}
\begin{tabular}{lclcccc}
\hline
Target           &    R.A.$^{a}$     &      Dec.$^{a}$     &
\multicolumn{3}{c}{$\mathrm{N}_{\mathrm{orb}} \times \mathrm{N}_{\mathrm{exp}} \times {\mathrm{t}_{\rm exp}} ^{b}$} & Prog. ID \\
                 &   [J2000]   &     [J2000]   & UVIS/F606W & IR/F105W & IR/F160W \\
\hline
Q0124+0044       & 01:24:03.78 &  +00:44:32.74 & $1\times 4\times 621$ & $1\times 4\times 653$ & $1\times 4\times 653$ & 14122 \\
Q0139--0824      & 01:39:01.41 & --08:24:44.05 & $1\times 4\times 622$ & $1\times 4\times 653$ & $1\times 4\times 653$ & 14122 \\
Q0310+0055       & 03:10:36.85 &  +00:55:21.66 &         --            & $1\times 4\times 653$ & $1\times 4\times 653$ & 14122 \\
Q0338--0005$^{c}$  & 03:38:54.78 & --00:05:21.01 & $2\times 4\times 627$ &          -- 		  &          --            & 12553 \\
Q0918+1636$^{c}$ & 09:18:26:16 &  +16:36:09.02 & $1\times 4\times 631$ & $1\times 4\times 653$ & $1\times 4\times 653$ & 12553 \\
Q1313+1441       & 13:13:41.20 &  +14:41:40.60 & $1\times 4\times 623$ & $1\times 4\times 653$ & $1\times 4\times 653$ & 14122 \\
Q2059--0528      & 20:59:22.43 & --05:28:42.78 & $1\times 4\times 622$ & $1\times 4\times 653$ & $1\times 4\times 653$ & 14122 \\
Q2222--0946$^{c}$& 22:22:56.11 & --09:46:36.29 & $1\times 4\times 629$ & $1\times 4\times 653$ & $1\times 4\times 653$ & 12553 \\
Q2239--2949      & 22:39:41.77 & --29:49:54.47 & $1\times 4\times 626$ & $1\times 4\times 653$ & $1\times 4\times 653$ & 14122 \\
Q2247--6015      & 22:47:08.93 & --60:15:45.30 & $1\times 4\times 646$ & $1\times 4\times 703$ & $1\times 4\times 703$ & 14122 \\
\hline
\end{tabular}
\begin{flushleft}
    $^{a}$ Right ascension and declination refer to the coordinates of
  the quasar.
    
    $^{b}$ $\mathrm{N}_{\mathrm{orb}}$ refers to the number of orbits,
  $\mathrm{N}_{\mathrm{exp}}$ to the number of exposures, and
  $\mathrm{t}_{\mathrm{exp}}$ to the exposure time of individual
  exposures in seconds.
    
    $^{c}$ Re-analysed \emph{HST} images of quasar fields. For
  Q0338--0005, the emission counterpart was originally detected with
  X-shooter spectroscopy \citep{Krogager2012}, whereas the \emph{HST}
  imaging data are presented in this work for the first time. For
  Q0918+1636, the emission counterpart was originally detected with
  X-shooter spectroscopy \citep{Fynbo2011}, and continuum emission was
  later detected with \emph{HST} imaging by \citet{Fynbo2013}. For
  Q2222--0946, the emission counterpart was originally detected with
  X-shooter spectroscopy \citep{Fynbo2010}, and continuum emission was
  later detected with \emph{HST} imaging by \citet{Krogager2013}.
\end{flushleft}
\end{table*}

\section{Observations and data reduction}
\label{sec:obs}

\subsection{Sample selection} 
\label{subsec:sampleselection}

The targets for our \emph{HST} observations were selected from
previously identified galaxy counterparts of strong HI absorbers at
$z\gtrsim2$ with spectroscopically confirmed redshifts based on
emission-lines. We do not include `proximate' absorbers ($\Delta
v_{\mathrm{QSO-abs}} < 5000~\mathrm{km}~\mathrm{s}^{-1}$). To optimise
our observational campaign, we pre-selected systems with high
metallicity ($\gtrsim 10$\% Solar) as these are more likely to have
luminous counterparts, thereby enabling a successful characterisation
of the emission. This selection strategy is an extension of the
successful X-shooter campaign targeting metal-rich DLAs (for details,
see \citealt{Fynbo2010} and \citealt{Krogager2017}). In total, we
selected ten quasar fields, with intervening absorber-galaxy pairs
that meet our metallicity and redshift criteria.
    
The \emph{HST} observations for the sample of ten is composed of new
and re-analysed imaging data of the quasar fields (see
Sect. \ref{subsec:hstdata}). We emphasise that all of the presented
\emph{HST} observations were carried out using the same observational
setup and strategy, which allows us to obtain homogeneous results. The
full sample analysed in this work is presented in
Table~\ref{tab:obslog}.

\begin{table*}
\caption{Literature data for our sample of strong \ion{H}{i} absorbers and their host galaxies.}
\label{tab:abschar}
\begin{tabular}{clccccccc}
\hline
Target      & $z_{\mathrm{QSO}}$ & $z_{\mathrm{abs}}$ & \logNHIcm & \multicolumn{2}{c}{$[\mathrm{M/H}]_{\mathrm{abs}} ^{\star}$} & $\theta$ & $b$ & SFR \\ 
 &  &  &   & Tracer &   & [arcsec] & [kpc] & [M$_{\odot}$~yr$^{-1}$]\\
\hline
Q0124$+$0044 & 3.84 & 2.2618$^{(j)}$ & $20.70\pm0.15^{(j)}$      & Zn\,\textsc{ii}    & $-0.67\pm0.16^{(j,s)}$ & $1.3^{(k)}$ & $10.9^{(k)}$ & ${>0.1_{\rm Ly\alpha}}^{(k)}$ \\
Q0139--0824  & 3.01 & 2.6773$^{(f)}$ & $20.70\pm0.15^{(f)}$     & Si\,\textsc{ii} & $-1.2\pm0.2^{(k)}$ & $1.6^{(k)}$ & $13.0^{(k)}$ & ${>0.7_{\rm Ly\alpha}}^{(k)}$ \\
Q0310$+$0055 & 3.78 & 3.1150$^{(a)}$ & $20.05\pm0.05^{(a)}$      & $-$ &    $-$ &   3.8$^{(a)}$ & 29.6$^{(a)}$ & ${>0.54_{\rm Ly\alpha}}^{(a)}$ \\
Q0338--0005  & 3.05 & 2.2298$^{(n)}$ & $21.09\pm0.10^{(n)}$      & Si\,\textsc{ii}  & $-1.37\pm0.06^{(e,s)}$ & 0.49$^{(o)}$ & 4.1$^{(o)}$ & ${>0.3_{\rm Ly\alpha}}^{(e)}$ \\
Q0918$+$1636 & 3.09 & 2.5832$^{(q)}$ & $20.96\pm0.05^{(q)}$    & Zn\,\textsc{ii}  & $-0.19\pm0.05^{(p,s)}$ & 2.0$^{(r)}$ & 16.4$^{(r)}$ & ${8\pm3_{\rm H\alpha}}^{(r)}$ \\
Q1313+1441   & 1.89 & 1.7941$^{(e)}$ & $21.30\pm0.10^{(e)}$      & Zn\,\textsc{ii}  & $-0.86\pm0.14^{(e,s)}$ & 1.3$^{(e)}$ & 11.3$^{(e)}$ & ${>0.3_{\rm Ly\alpha}}^{(e)}$ \\
Q2059--0528  & 2.54 & 2.2101$^{(b)}$  & $21.00\pm0.05^{(b)}$    & Zn\,\textsc{ii} & $-0.96\pm 0.06^{(b)}$ & $<$0.8$^{(b)}$ & $<6.3^{(b)}$ & ${0.2_{\rm Ly\alpha}}^{(b, e)} < \mathrm{SFR} < {1.4_{\rm H\alpha}}^{(l)}$\\
Q2222--0946  & 2.93 & 2.35409$^{(n)}$ & $20.65\pm0.05^{(p)}$    & Zn\,\textsc{ii}  & $-0.53\pm0.07^{(p)}$ & 0.80$^{(p,s)}$ & 6.7$^{(p)}$ & ${13\pm1_{\rm H\alpha}}^{(e)}$ \\
Q2239--2949  & 2.10 & 1.8250$^{(h,i)}$ & $19.84\pm0.14^{(g)}$  & Si\,\textsc{ii} & $>-0.67\pm 0.15^{(g)}$ & 2.4$^{(g)}$ & 20.8$^{(g)}$ & ${>0.07\pm0.01_{\rm Ly\alpha}}^{(g)}$ \\
Q2247--6015  & 3.01 & 2.3288$^{(c,d)}$ & $20.62\pm0.05^{(c)}$  & Zn\,\textsc{ii} & $-0.91\pm0.05^{(m,s)}$ & 3.1$^{(c,m)}$ & 26$^{(c,m)}$ & $33^{+40}_{-11}~ _{\rm H\alpha}^{~~(m)}$ \\
\hline
\end{tabular}
\begin{flushleft}
The projected separations (impact parameters) between the background
quasar and foreground galaxy are given as angular separations in
arcsec ($\theta$) and physical separations in kpc ($b$). All star
formation rates (SFRs) are reported for a Chabrier IMF. Sub-scripts on
SFRs refer to the spectral line from which the SFR was inferred. For
Q\,0124+0044 and Q\,0139--0824 the impact parameters and SFRs refer to
the new spectroscopic measurements published in this work
(Sect. \ref{subsubsec:Q0124+0044} and
Sect. \ref{subsubsec:Q0139-0824}, respectively). For Q\,0139--0824,
the metallicity refers to the value obtained from the absorption
analysis in this work (Appendix~\ref{subsec:absanalysis}).

$^{\star}$ All metallicities are given with respect to the Solar
reference as summarised by \citet[][their Table 1]{DeCia2016}.

{\bf References:}
   $(a)$ \citet{Kashikawa2014}, $(b)$ \citet{Hartoog2015},
   $(c)$ \citet{Bouche2012}, $(d)$ \citet{Lopez2002},
   $(e)$ \citet{Krogager2017}, $(f)$ \citet{Wolfe2008},
   $(g)$ \citet{Zafar2017}, $(h)$ \citet{Cappetta2010},
   $(i)$ \citet{Zafar2013}, $(j)$ \citet{Berg2016},
   $(k)$ This work, $(l)$ \citet{Peroux2012}, $(m)$ \citet{Bouche2013},
   $(n)$ \citet{Bashir2019}, $(o)$ \citet{Krogager2012},
   $(p)$ \citet{Fynbo2010}, $(q)$ \citet{Fynbo2011},
   $(r)$ \citet{Fynbo2013}, $(s)$ \citet{Moller2020}.
\end{flushleft}

\end{table*}

\subsection{HST data}
\label{subsec:hstdata}
We have acquired \emph{HST} broad-band images using the Wide Field
Camera 3 (WFC3) in three filters: F606W, F105W and F160W.  Seven of
the fields are new observations (ID 14122, PI: Christensen) and three
fields are re-analysed (ID 12553, PI: Fynbo), of which two were
published as single-object case-studies prior to this work
(Q0918+1636; \citealt{Fynbo2013} and Q2222-0946;
\citealt{Krogager2013}).  Given the redshifts of our targets, this
setup allows us to constrain the stellar continuum emission around the
rest-frame Balmer break, and thereby to constrain the stellar mass.
Each target was imaged during one orbit per filter (except
Q\,0338--0005, which was observed during two orbits in F606W only, and
Q\,0310+0055, which was observed only in the near-infrared filters),
subdivided into four exposures of equal lengths.  We used a standard
four-point dither pattern\footnote{WFC3-UVIS-DITHER-BOX for the F606W
  filter and WFC3-IR-DITHER-BOX-MIN for the F105W and F160W filters.}
observing strategy designed to provide an enhanced sub-pixel sampling
of the point-spread function (PSF).

The positions (impact parameter and position angle) of the emission
counterparts relative to the background quasars are known from
previous spectroscopic observations.  This allowed us to optimise the
orientation of the diffraction spikes of the PSF and to avoid detector
bleeding effects.  A log of the observations is provided in
Table~\ref{tab:obslog}.

To reduce individual exposures we use the official {\tt AstroDrizzle}
processing pipeline based on the Python package {\tt
  DrizzlePac}.\footnote{{\tt DrizzlePac} is a software product of the
  Space Telescope Science Institute (STScI).} For a detailed
description of {\tt AstroDrizzle}, we refer to the {\tt DrizzlePac}
documentation. The {\tt AstroDrizzle} procedure performs sky
subtraction and cosmic ray rejection before aligning and combining
individual exposures. For each filter, the individual exposures are
reconstructed on a sub-sampled pixel grid leveraging the sub-pixel
offsets used in the dither patterns to provide higher spatial sampling
in the final, combined images. For this purpose, we use a fixed `pixel
fraction' of 0.7 for all filters and a final pixel scale of
0.024~arcsec per pixel for the F606W observations and 0.067~arcsec per
pixel for the F105W and F160W observations. We use the same image
combination parameters for all targets, which allows us to construct
empirical, non-parametric PSFs based on the quasars themselves (see
Sect. \ref{subsec:psfsub}).

\subsection{Literature data}
\label{subsec:archivaldata}
In this Section, we review the relevant literature data compiled for
our sample.  In Table~\ref{tab:abschar}, we have collected the
previously published constraints regarding the spectroscopic
identifications of counterpart locations and limits on the derived
star formation rates (SFRs). For completeness, we also provide the
neutral hydrogen column densities and absorption-derived
metallicities.  More details for the individual absorbers are
presented below.

For consistency, we standardise all star formation rates to a Chabrier
IMF.  Literature values based on a Salpeter IMF are converted to their
Chabrier IMF equivalents by applying a downwards correction factor of
1.8.  For objects solely detected in Ly$\alpha$ emission, we have
implicitly assumed standard Case B recombination theory
($f_{\mathrm{Ly}\alpha} / f_{\mathrm{H}\alpha} = 8.7$) and escape
fractions of unity to convert the Ly$\alpha$ flux to a H$\alpha$-based
star formation rate following \citet{Kennicutt1998}.  Due to the
assumed escape fraction of 1.0, all SFR estimates based on \lya\ are
conservative lower limits. Furthermore, all measurements of line
fluxes based on slit spectroscopy, where the counterpart location was
not known in advance, provide lower limits to the total line flux due
to possible slit-loss.  We also provide Galactic extinction
corrections for the used photometric measurements based on
re-calibrated extinction maps by \citet{Schlafly2011}.

\subsubsection{Q\,0124+0044}
\label{subsubsec:Q0124+0044}
This quasar was observed as part of the large X-shooter legacy sample
`XQ-100' \citep[][Programme ID 189.A-0424]{Lopez2016}.  The X-shooter
slit was placed at a single position angle of $+130\degr$ east of
north that serendipitously contained emission from the galaxy
associated with the DLA at $z_{\mathrm{abs}}=2.2618$.  Remarkably, the
slit position was only $1\pm 1$ degree off from the true position
angle as measured on the \emph{HST} images presented in this work (see
Table \ref{tab:galfit}). This implies that slit-loss was minimal and
caused by slit-width, not by false angle.The emission counterpart was
discovered during the XQ-100 campaign by one of the authors of this
work (LC), but never reported in previous studies, as XQ-100 papers
focused on the absorption properties of the DLA in the 1-dimensional
(1D) quasar spectrum. Here we present an analysis of the 2-dimensional
(2D) X-shooter spectrum, and report on the emission associated with
the DLA counterpart.

\begin{figure}
\includegraphics[bb=80 518 1130 995,clip,width=0.48\textwidth]{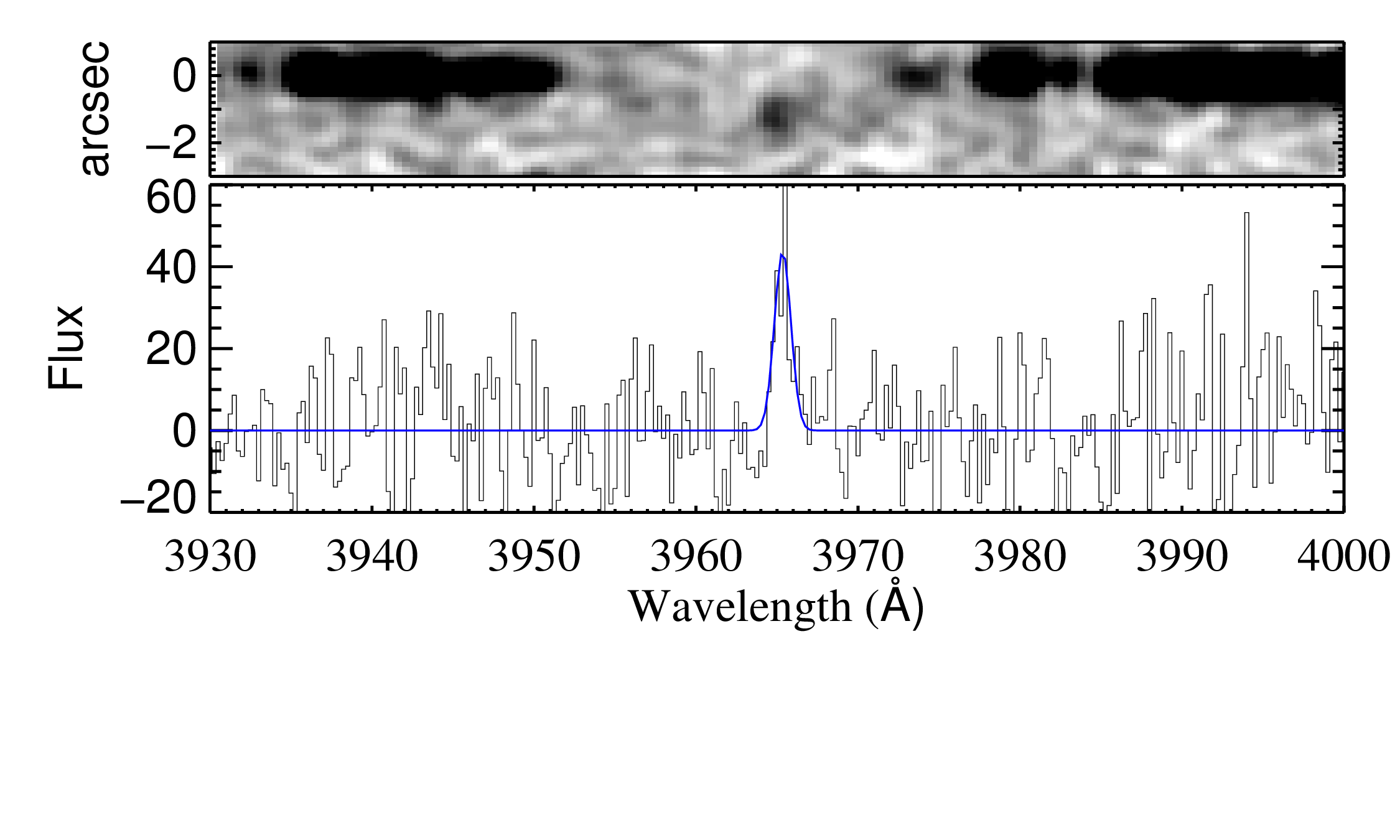}
\caption{Zoom-in of the Q\,0124+0044 X-shooter spectrum, centered on
  the Ly$\alpha$ line at $z = 2.2616$. The top panel shows the 2D
  spectrum, smoothed by a Gaussian filter with a FWHM of 3 pixels, and
  centered on the quasar trace. Ly$\alpha$ emission from the DLA
  counterpart can be seen at an impact parameter of $\sim$1.3~arcsec
  below the quasar. The bottom panel shows the 1D spectrum extracted
  at the position of the offset Ly$\alpha$ emission, employing a flux
  scale in units of $10^{-19}$~erg~s$^{-1}$~cm$^{-2}$~{\AA}$^{-1}$. A
  Gaussian profile was fitted to the emission line, and is overlayed
  on the 1D spectrum (in blue).
\label{fig:0124}}
\end{figure}

We detect Ly$\alpha$ emission at $z_{\rm em}=2.2616$, spatially offset
from the quasar by 1.3~arcsec (see Fig.~\ref{fig:0124}).  Integrating
over the line profile gives a line flux in \lya\ of $f({\rm Ly}\alpha)
= (5.8 \pm 0.3) \times 10^{-18}$~erg~s$^{-1}$~cm$^{-2}$ after
correcting for Galactic extinction, which yields a lower limit to the
SFR of $>0.1$~\MsunYr.

Extinction corrections are applied as follows for the F606W, F105W and
F160W filters: 0.071, 0.022 and 0.014~mag, respectively.

\subsubsection{Q\,0139--0824}
\label{subsubsec:Q0139-0824}

\begin{figure}
\includegraphics[bb=80 518 1130 995, width=0.49\textwidth]{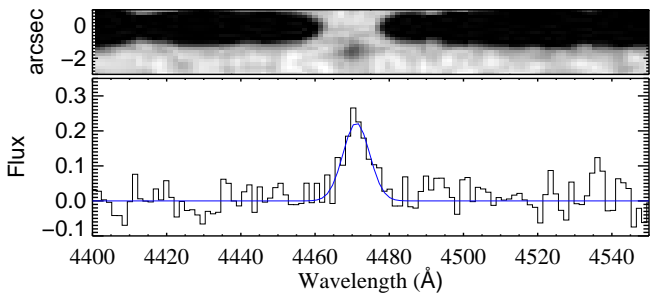}
\caption{Zoom-in of the Q\,0139--0824 FORS1 spectrum, centered on the
  Ly$\alpha$ line at $z = 2.6773$. The top panel shows the 2D
  spectrum, smoothed by a Gaussian filter with a FWHM of 2 pixels, and
  centered on the quasar trace. Ly$\alpha$ emission from the DLA
  counterpart can be seen at an impact parameter of $\sim$1.6~arcsec
  below the quasar. The bottom panel shows the 1D spectrum extracted
  at the position of the offset Ly$\alpha$ emission, employing a flux
  scale in units of $10^{-17}$~erg~s$^{-1}$~cm$^{-2}$~{\AA}$^{-1}$. A
  Gaussian profile was fitted to the emission line, and is overlayed
  on the 1D spectrum (in blue).}
    \label{fig:0139}
\end{figure}

The absorber towards Q\,0139--0824 was presented in \cite{Wolfe2008}.
A tentative detection of Ly$\alpha$ emission associated with the
absorber at $z_{\rm abs} = 2.6773$ was originally found with a low
signal in IFU data from VLT/VIMOS (Programme ID 077.A-0450, PI:
Christensen) but never published. To confirm the detection we
conducted deeper follow-up spectroscopic observations using the VLT
FORS spectrograph (Programme ID 081.A-0506, PI: Christensen). The
deeper data confirmed the detection of Ly$\alpha$, spatially offset by
$\sim$1.6~arcsec south-west of the quasar (see
Fig.~\ref{fig:0139}),and is published here for the first time. During
subsequent X-shooter observations aimed at identifying additional
emission lines (Programme ID 088.A-0378, PI: Christensen) a position
angle of $-98\degr$ east of north was used. No additional emission
lines at the DLA redshift were detected. However, given the redshift
of the DLA, the strongest rest-frame optical emission lines are
severely affected by telluric absorption.  We note that the slit
position angles used for spectroscopy do not correspond exactly to the
position angle of the galaxy measured directly in the \emph{HST}
images, since sufficient spatial information about the counterpart was
not available prior to the X-Shooter observations.

Integrating the Ly$\alpha$ line profile observed in the X-shooter
spectrum gives a line flux of $f({\rm Ly}\alpha) = (2.2 \pm 0.2)
\times 10^{-17}$~erg~s$^{-1}$~cm$^{-2}$ after correcting for Galactic
extinction, and yields a lower limit to the SFR of $>0.7$~\MsunYr.

In terms of origin and precision, the metallicity of the DLA towards
Q\,0139--0824 reported in the literature was rather uncertain. We
therefore performed a new analysis of the absorption system using the
X-shooter data. The new metallicity measurement is reported in
Table~\ref{tab:abschar} and the details of the absorption analysis are
presented in Appendix~\ref{subsec:absanalysis}.

Extinction corrections are applied as follows for the F606W, F105W and
F160W filters: 0.068, 0.022 and 0.014~mag, respectively.

\subsubsection{Q\,0310+0055}
The emission counterpart of the absorber towards Q\,0310+0055 
was detected using long-slit spectra from Subaru/FOCAS
\citep{Kashikawa2014}.

To accommodate the Subaru/FOCAS narrow-band detection
($\mathrm{mag}_{\mathrm{NB502}} = 25.46\pm 0.13$;
\citealt{Kashikawa2014}) in our stellar mass analysis (see
Sect.~\ref{subsec:SED}), we first remove the contribution from
Ly$\alpha$ line emission. Based on the reported Ly$\alpha$ luminosity
of $L_{\mathrm{Ly}\alpha} = (1.07 \pm 0.14) \times
10^{42}~\mathrm{erg~s}^{-1}$, and assuming a flat continuum shape over
the NB502 bandpass, with a filter width of
60~\AA \footnote{\url{https://www.naoj.org/Observing/Instruments/FOCAS/camera/filters.html}},
we obtain a NB continuum magnitude of $\mathrm{NB502}_{\mathrm{cont.}}
= 26.8\pm 0.6$.  For the SED fits using {\tt LePhare} (see
Sect.~\ref{subsec:SED}), we construct a simple tophat narrow-band
filter transmission curve with a central wavelength $\lambda_c =
5025$~{\AA} and a width of 60~{\AA}.  We also include their $B$- and
$V$-band non-detections (25.32 and 25.50 magnitudes, respectively)
reported as $3\sigma$ limits in 2 arcsec apertures at the position of
the narrow-band detection.  The observed Ly$\alpha$ line luminosity by
\citet{Kashikawa2014} corresponds to a lower limit to the SFR of
$>0.54$~\MsunYr.

All magnitudes (measurements and limits) were corrected for Galactic
extinction using the following corrections for the $B$, NB502, $V$,
F105W, and F160W filters: 0.414, 0.343, 0.305, 0.082, 0.052~mag,
respectively.

\subsubsection{Q\,0338--0005}
The emission counterpart of the DLA towards Q\,0338--0005 was
originally detected by \citet{Krogager2012} as part of the
VLT/X-shooter campaign, and was confirmed in an archival VLT/UVES
spectrum by \citet{Bashir2019}. \citet{Krogager2012} report the
detection of Ly$\alpha$ emission at an impact parameter of 0.5~arcsec
with a position angle of $-58\degr$ east of
north. \citet{Krogager2017} report a lower limit to the SFR of
$>0.3$~\MsunYr\ based on the line flux of Ly$\alpha$ as measured in
the long-slit spectra taken during the X-shooter campaign.

The Galactic extinction is rather high for this sightline, with an
extinction correction in the F606W filter of 0.207~mag.

\subsubsection{Q\,0918\,+1636}
There are two DLAs towards Q\,0918+1636. The detection of the emission
counterpart of the $z_{\rm abs} = 2.583$ DLA was reported by
\citet{Fynbo2011} and \citet{Fynbo2013} as part of the VLT/X-Shooter
campaign. The authors detect emission lines from [\ion{O}{ii}],
[\ion{O}{iii}], H$\alpha$ and H$\beta$ associated with the DLA at
$z_{\rm abs} = 2.583$ and infer a star formation rate of
8~\MsunYr\ based on the observed H$\alpha$ line flux.

Using \emph{HST} data, \citet{Fynbo2013} detect the continuum emission
of the DLA at $z_{\mathrm{abs}} = 2.583$ at an impact parameter of
1.98~arcsec from the quasar at a position angle of $-115\degr$ east of
north, and obtain a stellar mass of $\log ({\rm M}_{\star} / {\rm
  M_{\odot}}) = 10.1_{-0.1}^{+0.2}$.

In addition to the re-reduced \emph{HST} images and their associated
magnitudes, we also include the Galactic extinction corrected SDSS
$u$, $g$, and K$_s$-band observations from the Nordic Optical
Telescope NOTCam \citep{Abbott2000} reported by \citet{Fynbo2013}. For
the F606W, F105W and F160W magnitudes we adopt extinction corrections
of 0.058, 0.018 and 0.012~mag, respectively.

\citet{Fynbo2013} furthermore detect emission lines from Ly$\alpha$
and [\ion{O}{iii}] $\lambda$5007 associated with the second absorber
at $z_{\rm abs} = 2.412$ at a very small impact parameter
$<0.25$~arcsec at a position angle of $\sim 130$ deg east of north.
We are not able to test for such an object in our \emph{HST} imaging
data, as detailed in Sect.~\ref{subsec:psfsub}, due to the strong
residuals of the quasar PSF subtraction at these small impact
parameters.

\subsubsection{Q\,1313+1441}
\label{subsubsec:Q1313+1441}

The emission counterpart of the absorber towards Q\,1313+1441 was
detected by \citet{Krogager2017} as part of the X-shooter campaign
\citep{Fynbo2010, Krogager2017}.  The detection is based on Ly$\alpha$
emission in the trough of the damped Ly$\alpha$ absorption profile at
an impact parameter of 1.3~arcsec.  The authors report detections in
two slits with different orientations yielding seemingly inconsistent
relative positions of the emission.  The quoted impact parameter
refers to the brighter of the two detections.

\citet{Krogager2017} suggest that the two detections may arise from
two different neighbouring galaxies. With the \emph{HST} data in hand,
we can revise this explanation. We do not detect multiple counterparts
around the quasar that would coincide with the detections reported by
\citet{Krogager2017}. Instead, we identify a single counterpart in the
near-infrared data with an extended, disturbed structure (see
Sect.~\ref{subsec:psfsub}).  Such an extended structure could explain
the appearance of emission in the two slits at orientations of +60 and
$-$60$\degr$ (east of north) used by the authors.

Extinction corrections for the F606W, F105W and F160W filters are
applied as follows: 0.051, 0.016 and 0.010~mag, respectively.

\subsubsection{Q\,2059--0528}
The emission counterpart of the DLA towards Q\,2059--0528 was detected
by \citet{Hartoog2015}, as part of the X-Shooter campaign
\citep{Fynbo2010, Krogager2017}.  The authors tentatively detect
Ly$\alpha$ emission in all three slit orientations (at $\sim 3 \sigma$
in individual slits).  By stacking all three observations,
\citet{Hartoog2015} obtain a robust detection in Ly$\alpha$ of $f_{\rm
  Ly\alpha} = (10.19\pm1.67) \times 10^{-18}$~erg~s$^{-1}$~cm$^{-2}$.
The Ly$\alpha$ line flux provides a lower limit to the SFR of
$>0.2$~\MsunYr.  The fact that emission is seen in all three slits
suggests a low impact parameter for the counterpart ($<$0.75~arcsec).

\citet{Peroux2012} used VLT/SINFONI IFU observations to search for
H$\alpha$ emission associated to the absorber. They report a
non-detection, which translates into an upper limit to the SFR of
$<1.4$~\MsunYr.

Extinction corrections for the F606W, F105W and F160W filters are
applied as follows: 0.104, 0.033 and 0.021~mag, respectively.

\subsubsection{Q\,2222\,--0946}
The counterpart of the absorber towards Q\,2222--0946 was detected by
\citet{Fynbo2010} as part of the VLT/X-shooter campaign. The authors
report a detection of the emission counterpart at an impact parameter
of 0.8~arcsec, at a predicted position angle of $\sim 40\degr$ based
on triangulation from the observations with different slit
orientations. Using \emph{HST} data, \cite{Krogager2013} confirm the
detection at an impact parameter of 0.74~arcsec.

Using deep VLT/X-shooter data obtained with the slit aligned towards
the emission counterpart, \citet{Krogager2013} detect emission from
Ly$\alpha$, [\ion{O}{ii}], [\ion{O}{iii}], H$\alpha$, and H$\beta$
lines.  Further data was obtained from Keck/Osiris
\citep{Jorgenson2014} and VLT/SINFONI
\citep{Peroux2012}. \citet{Krogager2013} report a SFR based on
H$\alpha$ of 13~\MsunYr\ and a stellar mass of $\log ({\rm M}_{\star}
/ {\rm M_{\odot}}) = 9.3\pm0.2$ based on SED fitting.

For the F606W, F105W and F160W magnitudes we adopt extinction
correction factors of 0.103, 0.032 and 0.021 magnitudes, respectively.

\subsubsection{Q\,2239--2949}
The emission counterpart of the absorber towards Q\,2239--2949 was
identified by \citet{Zafar2017}. The authors report a detection of
Ly$\alpha$ emission at $z_{\rm abs} = 1.825$ spatially offset by
2.4~arcsec from the quasar. The Ly$\alpha$ emission line flux of the
counterpart translates into a lower limit on the SFR of $>0.07\pm
0.01$~\MsunYr. The absorber is a sub-DLA and no ionisation correction
was applied to the reported metallicity, which should therefore be
interpreted with care.

We adopt the following extinction corrections for the F606W, F105W and
F160W filters: 0.043, 0.013, and 0.009~mag, respectively.

\subsubsection{Q\,2247--6015 (alternative name: HE\,2243--60)}
The absorber towards Q\,2247--6015 was first analysed by
\citet{Lopez2002} using data from VLT/UVES. \citet{Bouche2012,
  Bouche2013} conducted comprehensive followup observations of the
quasar field using VLT/SINFONI IFU data and detect the associated
counterpart in H$\alpha$ at an impact parameter of 3.1~arcsec.  The
authors report a dust-corrected H$\alpha$ star formation rate of
33~\MsunYr, which owing to the aperture-free IFU data is not affected
by slit-losses.

Extinction corrections for the F606W, F105W and F160W filters are
applied as follows: 0.047, 0.015 and 0.009~mag, respectively.

\section{Results}
\label{sec:results}

\begin{figure*}
\centering
\includegraphics[width=0.78\textwidth]{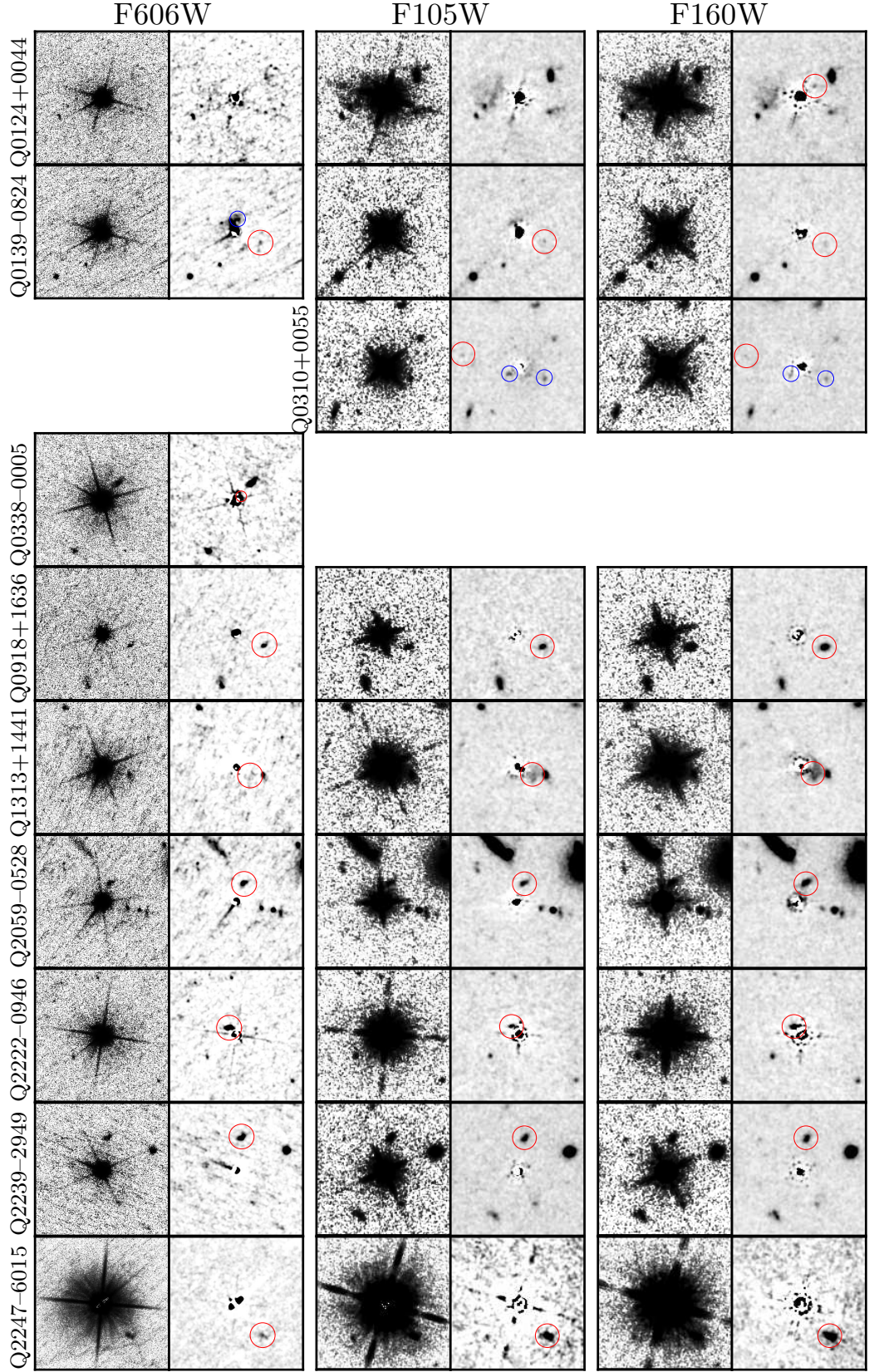}
\caption{Mosaic of 9$\times$9 arcsec cutouts centered on each quasar.
  Rows show individual targets. Each column of two sub-panels shows
  one \HST\ filter.  Each sub-panel displays the science image
  (\textit{left}) and quasar subtracted residual image
  (\textit{right}) with confirmed detections (red circles) and
  proximate candidates with no redshift information (blue
  circles). Each panel is aligned north up, east left. Science images
  are displayed with a histogram equalisation to enhance faint
  structures, and residual images are Gaussian smoothed with a fixed
  smoothing-length of 0.06~arcsec.}
\label{fig:PSFs}
\end{figure*}

Based on the \HST\ observations presented in
Sect.~\ref{subsec:hstdata}, we successfully detect continuum emission
for all targets in our sample in at least one filter (nine robust
detections; one tentative detection in the Q\,0338--0005 field, see
Fig.~\ref{fig:PSFs} and Fig.~\ref{fig:my_label}).  Eight of these
continuum detections are presented in this work for the first time
(the seven targets associated with Programme ID 14122 and the
previously unpublished Q\,0338--0005 observation associated with
Programme ID 12553).  We analyse the surface brightness distributions
of the detected continuum emission and provide measurements of the
half-light radii (see Sect.~\ref{subsec:galfit}).

For eight of the ten targets in our statistical sample (six associated
with Programme ID 14122 and the two re-analysed objects associated
with Programme ID 12553), we secure detections in more than one
filter, which enables us to estimate the SFR and stellar mass by
modelling the photometric spectral energy distributions (see
Sect.~\ref{subsec:SED}). Our analysis of the \HST\ data is presented
below.

\subsection{Subtraction of quasar point spread function}
\label{subsec:psfsub}

To detect the faint stellar continuum of the absorbing galaxy and to
search for objects hiding under the bright quasar PSF, we must isolate
the flux from the quasar and subtract it from each
image. Traditionally for \emph{HST} images, a synthetic PSF is either
created using {\tt TINYTIM}, or empirically modelled from bright,
unsaturated stars in the same exposure \citep[e.g.][]{Kulkarni2000,
  Krogager2013, Fynbo2013}. The former method is noiseless, can be
constructed for the position of the quasar on the detector, and
captures the profile of the outer PSF wings. However, the model is
limited by the details in its construction; by the accuracy of the
recorded telescope aberrations; and can produce unsatisfactory models
for saturated objects \citep[e.g.][]{Krogager2013}. The latter,
empirical method takes advantage of the high S/N of bright stars and
is observed simultaneously with the quasar, which mitigates temporal
differences. However, this method is sensitive to the position of the
bright star on the detector plane and aberrations of the
telescope. The empirical approach is often unable to model the
extended wings of the PSF as the outer regions are dominated by noise
in the sky background. Further limitations include the number of
suitable stars in the field as well as potential colour differences
between the object of interest (in our case, the quasar) and the stars
used to construct the PSF model \citep{Warren2001}.

Since our \emph{HST} programme targets multiple quasar fields with the
same observing strategy, and since the final data products are
combined with identical settings (see Sect.~\ref{subsec:hstdata}), the
aforementioned caveats can be mitigated by modelling the quasar PSF
(qPSF) from the targeted quasars themselves \citep[e.g.][]{Warren2001,
  Augustin2018}.  For each quasar ($i=1...10$) in a given band, we
construct empirical, non-parametric models of qPSF$_i$ from the
median-combined stack of all remaining qPSFs of the same
filter. Before combining the images, the individual images are
sub-sampled on a 4-times finer grid and re-centered to the PSF
centroid.

Finally, our empirically constructed qPSF models are resampled back to
the original binning and fitted to their respective quasar image in
order to subtract the quasar emission, see Fig.~\ref{fig:PSFs}.  After
subtracting the qPSF, we search the field for faint stellar continuum
emission associated with the foreground absorber at the locations
identified in previous spectroscopic observations (highlighted as red
circles in Fig.~\ref{fig:PSFs}).  Moreover, we can look for potential
candidates at smaller impact parameters down to a limit of $\sim
0.5$~arcsec where residuals from the qPSF subtraction start to
dominate (blue circles in Fig.~\ref{fig:PSFs}).

In the field of Q\,0139--0824, we identify a bright object in the
F606W band at low impact parameter (blue circle in
Fig. \ref{fig:PSFs}). This object clearly overlaps with the quasar
PSF, but is unidentified in the other bands. We therefore revisited
the quasar spectrum to see whether this object has an absorber
counterpart. The spectrum reveals a weak absorber at a redshift
$z_{\rm abs} = 2.233$, which we hypothesise is the counterpart of the
bright object. However, the absorption lines are too weak to be
compatible with a strong \ion{H}{i} absorber, and we therefore leave
it to be pursued in future work. Alternatively, the object may be part
of the quasar's local galactic environment.

For Q\,0310+0055, we note that our PSF subtracted images reveal two
objects at lower impact parameters (blue circles in
Fig.~\ref{fig:PSFs}).  However, \cite{Kashikawa2014} do not detect
\lya\ emission at the absorber redshift at lower impact parameters.
Since evidence for a physical connection is absent, we disregard them
in the remainder of this work, but note that the detections may be
either nearby galaxies that do not emit \lya; part of the local
environment of the quasar; or simply low-redshift interlopers.
Further spectroscopic investigation of the field is needed to reveal
the nature of these objects.

\begin{figure}
\centering
\scalebox{1}[1]{\includegraphics[width=0.38\textwidth]{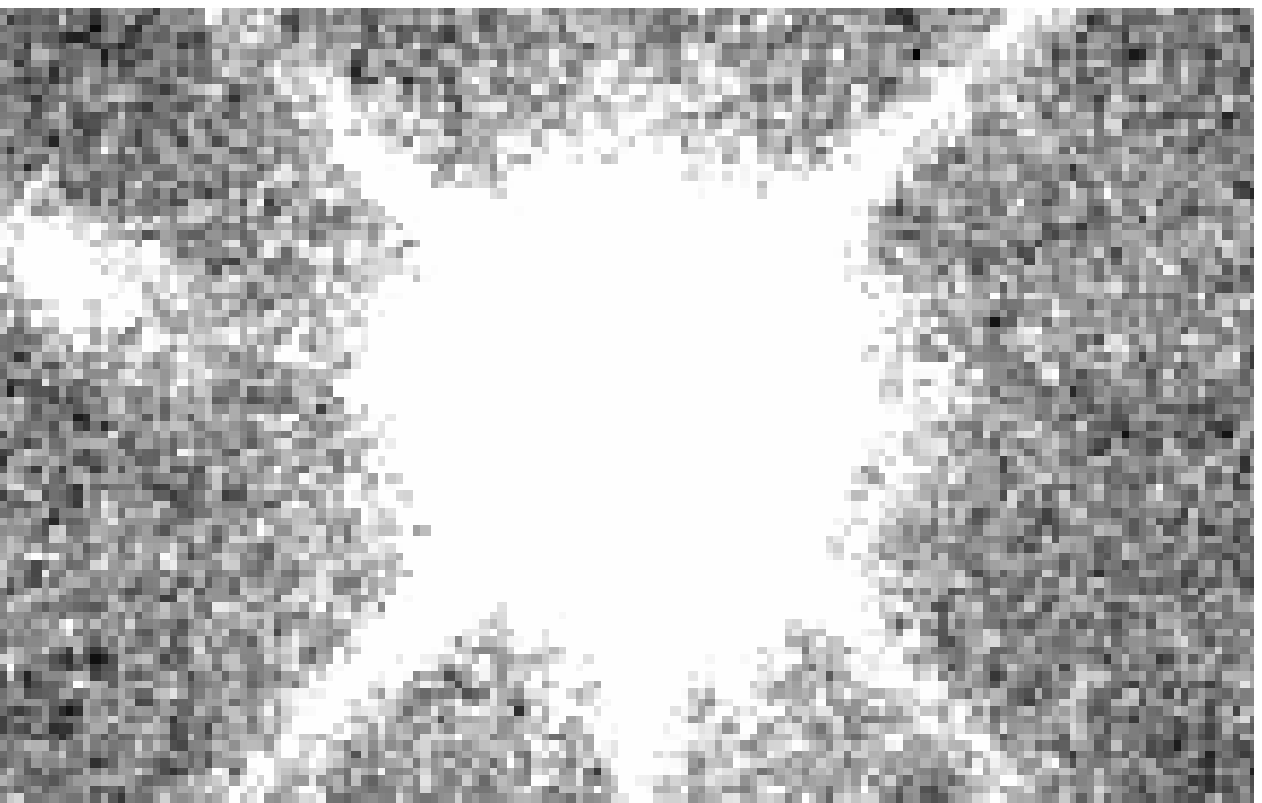}}

\vspace{1mm}
\scalebox{1}[1]{\includegraphics[width=0.38\textwidth]{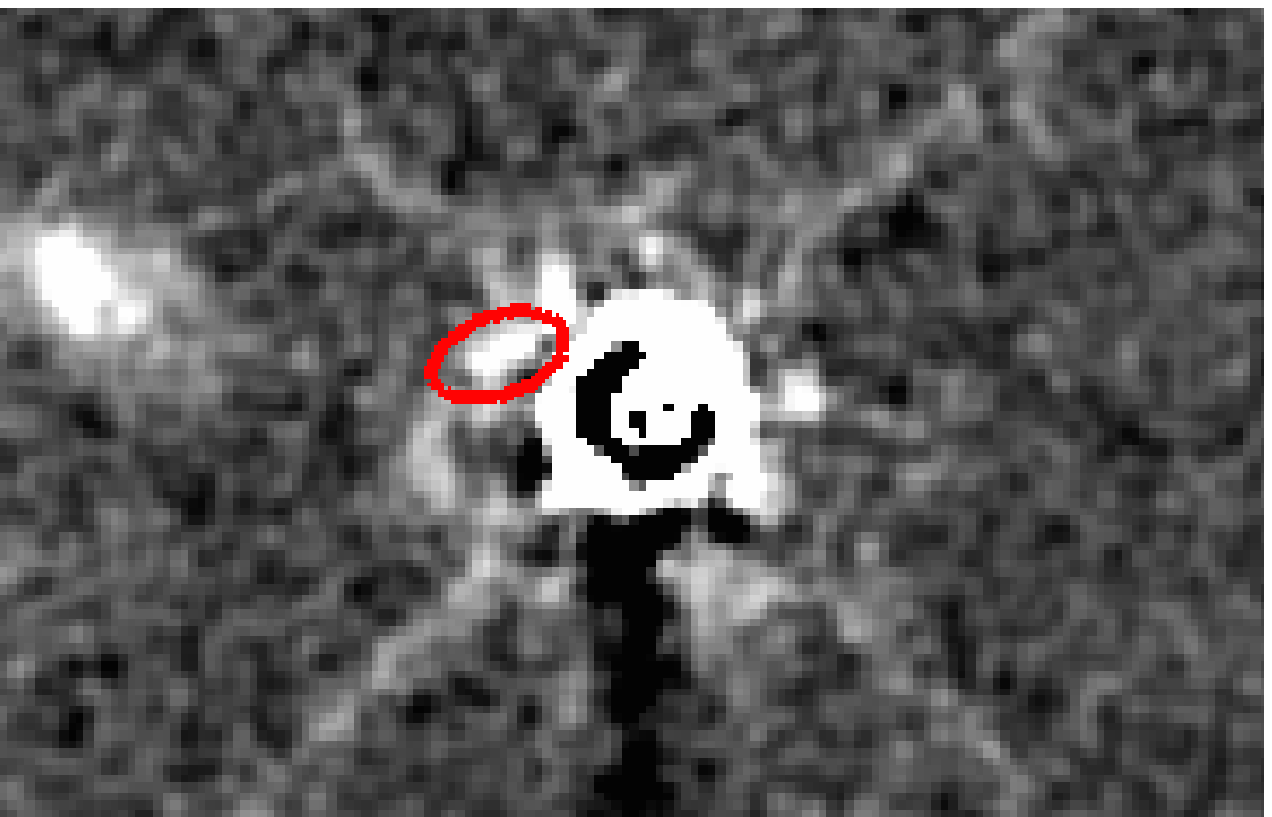}}

\vspace{1mm}
\scalebox{1}[1]{\includegraphics[width=0.38\textwidth]{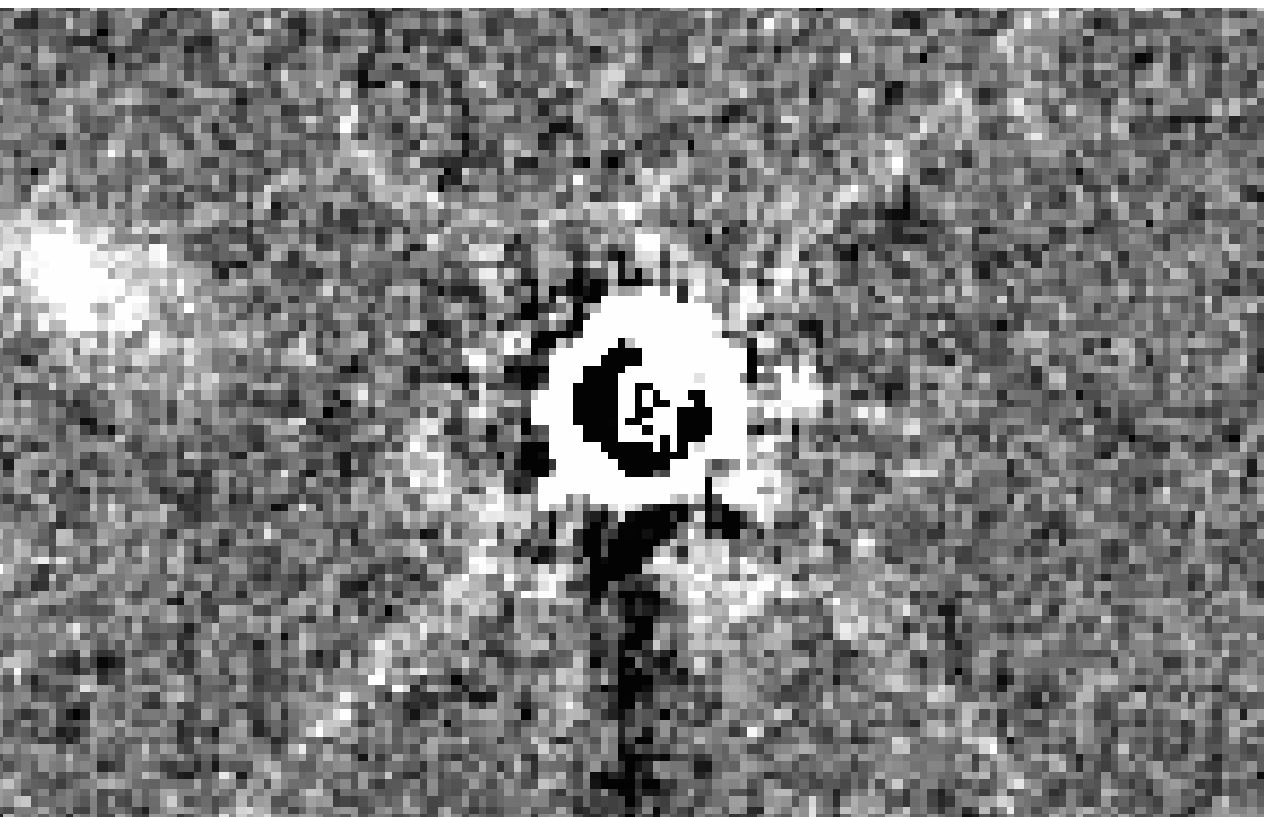}}
\caption{ Tentative detection of the counterpart associated with the
  DLA towards Q\,0338--0005. The panels show from top to bottom: a
  zoom-in of the quasar; the image after subtraction of the qPSF; and
  the residuals after simultaneously fitting the qPSF and the
  tentative counterpart.  The location of the tentative counterpart is
  shown by the red ellipse in the middle panel. Note that the panels
  display the native image axes.  }
\label{fig:my_label}
\end{figure}

In the field of Q\,0338--0005, we only obtain a tentative detection of
the counterpart observed in spectroscopy by \citet{Krogager2012}. The
tentative detection of the continuum emission of the counterpart is
shown in Fig.~\ref{fig:my_label}, located at an impact parameter of
$0.39\pm0.02$~arcsec from the quasar line of sight at a position angle
of $-55\pm2\degr$.  The impact parameter and position angle are
consistent with those reported by \citet{Krogager2017}. Nonetheless,
given the strong residuals of the PSF subtraction we only consider
this a tentative detection.

For Q\,2059--0528, the PSF-subtracted image reveals multiple objects,
all of which have larger impact parameters than the limit ($b <
0.75$~kpc) reported by \cite{Hartoog2015}. We note, however, that this
limit is based on the assumption that the individual Ly$\alpha$
signals in the three slits are detecting the same
counterpart. Relaxing this assumption, we note that their reported
detection in the $\mathrm{PA}= -60\degr$ slit is the only detection
formally above $3\sigma$ significance. We therefore identify the
bright object immediately north-west of the quasar as the most likely
source of the Ly$\alpha$ emission, and as a candidate counterpart to
the DLA. The impact parameter of 1.43~arcsec measured in our
\HST\ image is furthermore consistent with the location of the
emission in the spectrum with slit angle $\mathrm{PA}= -60\degr$ by
\citet{Hartoog2015}.  Such a configuration with several components
seen in emission appears similar to the DLA host galaxy system towards
Q\,2206--1958, shown to be in an active stage of merging
\citep{Moller2002, Weatherley2005}.

\subsection{Modelling surface brightness profiles}
\label{subsec:galfit}

Having identified the continuum emission counterparts, we use {\sc
  galfit} \citep{Peng2002} to model their surface brightness
profiles. Here, we do so by iteratively adding S\'ersic components
that are fitted simultaneously with the quasar point source until the
galaxy emission is fully captured. We set up {\sc galfit} to resample
the PSF back to its original binning during fitting.  {\sc galfit}
automatically takes the PSF convolution into account when performing
the fit.

{\sc galfit} has been used to derive structural parameters and
magnitudes for individual absorption-selected galaxies in the past
\citep[e.g.][]{Krogager2013,Fynbo2013,Augustin2018}. We here make an
effort to outline and emphasise certain aspects of the fitting
procedure in this work that differ from the standard case of
retrieving parameters for a galaxy on a flat sky background.

\begin{itemize}
    \item The brightness of the quasar causes the qPSF profile to
      extend far into the field. It is therefore essential to use a
      large PSF model, capable of subtracting the flux in the PSF
      wings. If the PSF does not account for this, {\sc galfit} will
      overestimate the sky background.
    \item The large PSF model necessitates a large fitting region. We
      found that a fitting region of 800$\times$800 pixels for a pixel
      scale of 0.067~arcsec per pixel was needed to ensure robust
      quasar magnitudes matching known SDSS optical photometry.
    \item The S\'ersic function associates higher light concentrations
      with larger extended wings. Any over/under estimation of either
      the sky background and/or the quasar PSF wing will therefore be
      compensated for by (wrongfully) adjusting the S\'ersic
      index. This will minimise the $\chi ^2$ statistic at the expense
      of unrealistically large concentrations. It is therefore
      essential to fit the galaxy and quasar simultaneously, and
      determine the background independently. This is particularly
      important as our objects lie at small impact parameters.  Given
      the high contrast between the quasar and the galaxy brightness,
      a small change in the quasar magnitude may generate large
      differences in final galaxy parameter estimates.
\end{itemize}

We therefore fix the sky value to independent measurements determined
from the mean of the pixel counts in sky regions free of sources and
hot pixels. The zero-point (ZP) for each filter was calculated with
the PHOTPLAM and the PHOTFLAM FITS header keywords, giving values of
ZP$_{\rm F606W} = 26.104$; ZP$_{\rm F105W} = 26.270$; and ZP$_{\rm
  F160W} = 25.947$. Finally, we pass the science image to {\sc galfit}
in the recommended units of counts. This is particularly relevant for
our science case, as the brightness contrast between the quasar and
the galaxy causes pixel values to span a large dynamic range.

Adopting a single S\'ersic component does not capture the clumpy light
distribution observed in many of the objects, but results in large
residuals. This behaviour is also reflected in the {\sc galfit}
solution, which under such conditions (or for faint objects) becomes
unstable to perturbations in initial parameter values. Iteratively
adding S\'ersic components, which collectively capture the effective
light distribution of the source, stabilises the parameter range
within physically acceptable S\'ersic indices $(0.5 \leq n \leq 8)$
and effective semi-major axes ($0.2 \leq a_e~/~\mathrm{pixels} \leq
20$), although the distribution of the parameters remains sensitive to
the initial guesses, giving near-identical $\chi^2$-statistics for
different combinations.

For Q\,1313+1441, the fit to the extended and disturbed structure
identified as the emission counterpart requires the introduction of an
intermediate S\'ersic profile to converge on a satisfactory model. The
{\sc galfit} solution suggests a bright component $\sim0.7$ arcsec
from the quasar, but at this proximity it is sensitive to quasar
subtraction residuals, and it lies a factor of two off the
triangulated position (see Sect. \ref{subsubsec:Q1313+1441}). We
therefore choose to disregard this component when reporting the
results, and emphasise that further work is needed to establish the
nature of the bright signal detected in both the IR bands.

The final fits suggest that $\sim 40\%$ of absorption-selected
galaxies at redshift $z\sim 2-3$ require multiple S\'ersic profiles to
mitigate apparent systematic residuals and accurately capture the
light distribution (for individual number of S\'ersic components
employed in each fit, see Table~\ref{tab:galfit}).  We note that the
number of S\'ersic profiles employed for each host galaxy should be
considered lower limits, as they correlate with spatial
resolution. However, the fact that $40\%$ reveal multiple star-forming
clumps at a drizzled F160W spatial resolution of 0.067~arcsec per
pixel at $z\sim 2-3$ agrees with the general trend of clumpy
morphologies of high-redshift galaxies \citep{Livermore2015}.

\subsubsection{Non-parametric size measurements}

Previous studies, often motivated by observational considerations,
chose to report structural parameters and morphology based on the band
with the highest spatial resolution. Here, we attempt a physically
motivated approach, and report morphologies based on the reddest
(F160W) band in order to capture the main stellar component. We then
fix the morphology in the remaining bands to that derived in F160W,
appropriately scaled and rotated to the resolution, pixel sampling and
orientation of individual frames.

To systematically analyse and compare objects fitted with a single
S\'ersic profile to those that require multiple components we adopt a
conservative approach, and report non-parametric half-light radii
($r_{1/2}$) calculated from growth curves originating at the centroid
of each galaxy model together with its total magnitudes (summed over
individual S\'ersic components).  For $r_{1/2}$ the final uncertainty
reflects an inverse variance weighted sum of the relative
uncertainties from each component contributing to the model. We
estimate magnitude errors from the flux of the modelled light
distribution and flux errors measured directly in the quasar and
galaxy subtracted residual image as $\Delta F=\sqrt{\sum_{i \in A}
  \sigma^2_{\mathrm{tot},i}}$, where $A$ is the non-masked pixels in a
$5 \times r_{1/2}$ circular aperture at the position of the centroid
of the galaxy model, and $\sigma_{\mathrm{tot},i}$ is the residual
flux in pixel $i$. These results are recorded in
Table~\ref{tab:galfit}.

\subsubsection{Comparison to literature measurements}

Independent {\sc galfit}-based sizes were reported for
the sources in the fields of Q\,2222--0946 
\citep[][both based on the F606W filter]{Krogager2013, Augustin2018}
and Q\,0918+1636 \citep[][based on the F160W filter]{Fynbo2013}.
In the former field, \citet{Krogager2013} report an effective semi-major
axis of $a_e = 1.12 \pm 0.06$~kpc, and \citet{Augustin2018} find a
corresponding value of $a_e = 1.5 \pm 0.1$~kpc. To compare these size
measurements with our growth curve based $r_{1/2}$, we convert the
reported $a_e$ to circularised radii as
$r_{\text{circ}} = a_e \times \sqrt{b/a}$,
where $b/a$ refers to the semi-major-to-minor axis-ratio.
For $b/a = 0.17 \pm 0.02$ \citep[][]{Krogager2013} we obtain
$r_{\text{circ}} = 0.46\pm 0.03$~kpc (consistent with our $r_{1/2}$ measurement within $1.4\sigma$). \citet{Augustin2018} report an axis ratio of $b/a = 0.38 \pm 0.06$, which translates to $r_{\text{circ}} = 0.9\pm 0.1$~kpc (indicating a $\sim 3.6\sigma$ tension with our value).

For the counterpart of the $z=2.583$ DLA towards Q\,0918+1636,
\citet{Fynbo2013} report values of $a_e = 1.4 \pm 0.1$~kpc and $b/a =
0.4 \pm 0.1$ which translates into $r_{\text{circ}} = 0.9 \pm
0.2~\mathrm{kpc}$, perfectly consistent with our $r_{1/2}$
measurement. These results demonstrate that our generalised method to
retrieve non-parametric half-light radii is robust and provides
consistent results with those from the literature.  The $\sim
3.6\sigma$ tension between our $r_{1/2}$ and the circularised
effective radius reported by \citet[][]{Augustin2018} cannot purely be
attributed to the analysis being performed in different filters
(theirs using F606W; ours using F160W) and therefore sampling
different underlying stellar populations. Our size measurement is
consistent with the value obtained by \citet[][]{Krogager2013}, based
on their analysis of the F606W filter image.  Nor can the discrepancy
be explained by methodological differences, as our $r_{1/2}$
measurements agree remarkably well with the $r_{\text{circ}}$ of two
separate studies, conducted in two independent quasar fields
\citep[][]{Fynbo2013, Krogager2013}.

\begin{table*}
\caption{Results of {\sc galfit} modelling of absorber counterparts.}
\label{tab:galfit}
\begin{tabular}{clccccccc}
\hline
Target & N$_{\rm S}$	& P.A. & $\theta$ & $b$ & $r_{1/2}$  & mag$_{\mathrm{F606W}}$ & mag$_{\mathrm{F105W}}$ & mag$_{\mathrm{F160W}}$ \\
 & & [deg]		& [arcsec] & [kpc] &  [kpc] &  [AB] & [AB] & [AB] \\
\hline
Q0124$+$0044 & $1$ & $311\pm1$ & $1.25\pm0.04$ & $10.5\pm0.3$ & $0.4\pm3.1$  & -- & -- & $26.9\pm0.2$ \\ 
Q0139$-$0824 & $1$ & $244.6\pm0.7$ & $1.83\pm0.03$ & $14.8\pm0.2$ & $0.4\pm1.5$  & $27.2\pm0.1$ & $27.3\pm0.2$ & $26.8\pm0.1$ \\ 
Q0310$+$0055 & $1$ & $77.9\pm0.5$ & $3.77\pm0.04$ & $29.4\pm0.3$ & $0.5\pm0.9$  & -- & $26.9\pm0.1$ & $26.7\pm0.1$ \\ 
Q0338$-$0005 & $1$ & $305\pm2$ & $0.39\pm0.02$ & $3.3\pm0.2$ & $1.5\pm0.1 ^{a}$  & $25.7\pm0.1 ^{a}$ & -- & -- \\ 
Q0918$+$1636 & $1$ & $245.78\pm0.06$ & $2.00\pm 0.01$ & $16.37\pm0.02$ & $0.81\pm0.03$  & $25.65\pm0.06$ & $24.53\pm0.03$ & $23.66\pm0.01$ \\
Q1313$+$1441 & $2$ & $241\pm2$ & $1.20\pm0.04$ & $10.3\pm0.4$ & $2.2\pm0.7$ &  $26.7\pm 0.4$ & $26.5\pm0.3$ & $25.7\pm0.2$ \\ 
Q2059$-$0528 & 2 & $334.8\pm0.3$ & $1.43\pm0.01$ & $12.05\pm0.09$ & $1.7\pm0.3$  & $25.23\pm0.08$ & $24.71\pm0.08$ & $23.74\pm0.06$ \\ 
Q2222$-$0946 & $1$ & $41.1\pm0.2$ & $0.74\pm 0.01$ & $6.17\pm0.02$ & $0.52\pm0.03$  & $24.32\pm0.03$ & $24.59\pm0.09$ & $23.63\pm0.06$  \\ 
Q2239$-$2949 & 3 & $348.8\pm0.7$ & $2.31\pm0.03$ & $19.8\pm0.3$ & $1.5\pm0.4$ & $25.00\pm0.06$ & $24.39\pm0.03$ & $24.05\pm0.03$ \\ 
Q2247$-$6015 & $2$ & $222.3\pm0.3$ & $2.99\pm0.02$ & $25.0\pm0.2$ & $2.6\pm0.3$  & $23.48\pm0.06$ & $23.33\pm0.07$ & $22.65\pm0.04$ \\ 
\hline
\end{tabular}

\begin{flushleft}
    N$_{\mathrm{S}}$ refers to the number of S\'ersic components used
    in the fit.  The position angle (P.A.) of the galaxy position
    relative to that of the quasar is measured in degrees east of
    north.  Magnitudes are tabulated without correction for Galactic
    extinction.
    
    $^{a}$ The fit of the tentative counterpart of Q0338--0005
    required highly fine-tuned parameters. The half-light radius and
    the associated magnitude should therefore be treated with caution.
\end{flushleft}
\end{table*}

\begin{table}
\caption{Results of SED modelling of absorber counterparts.}
\label{tab:SFR}
\begin{tabular}{lccc}
\hline
Target       & $E_{B-V}$ & $\log (\text{SFR} / \mathrm{M}_{\odot} ~ \mathrm{yr}^{-1})$ & $\log (\mathrm{M}_{\star} / \mathrm{M}_{\odot})$ \\
\hline
Q0124$+$0044 &     --    &       --            &      --              \\
Q0139$-$0824 &    0.00   & $0.0\pm0.3$        & $8.2\pm 0.2$         \\
Q0310$+$0055 &    0.00   & $0.4\pm0.3$         & $8.6\pm 0.3$         \\
Q0338$-$0005 &     --    &      --             &      --              \\
Q0918$+$1636 &    0.25   & $1.3\,^{+0.1}_{-0.2}$ & $10.2\,^{+0.1}_{-0.2}$ \\
Q1313$+$1441 &    0.00   & $0.0\pm0.3$         & $8.6\,^{+0.4}_{-0.6}$  \\
Q2059$-$0528 &    0.20   & $1.2\,^{+0.3}_{-0.4}$ & $9.7\,^{+0.6}_{-0.5}$  \\
Q2222$-$0946 &    0.00   & $0.9\,^{+0.2}_{-0.1}$ & $9.2\pm 0.1$  \\
Q2239$-$2949 &    0.35   & $1.7\pm 0.2$        & $9.0\pm 0.1$         \\
Q2247$-$6015 &    0.10   & $1.6\pm 0.3$        & $9.8\,^{+0.7}_{-0.6}$ \\
\hline
\end{tabular}

\begin{flushleft}
    Star formation rates and stellar masses are reported as median
    values with $1\sigma$ uncertainties based on the 16th- and 84th
    percentiles, determined using the built-in maximum likelihood
    routine of {\sc LePhare}.
\end{flushleft}

\end{table}

\begin{figure*}
\centering
    \includegraphics[width=0.38\textwidth]{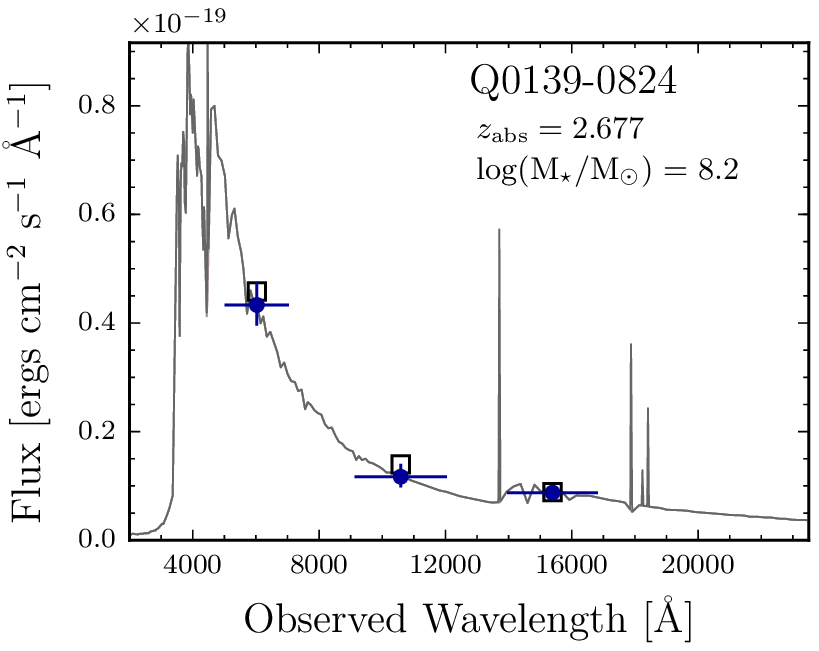}
    \includegraphics[width=0.38\textwidth]{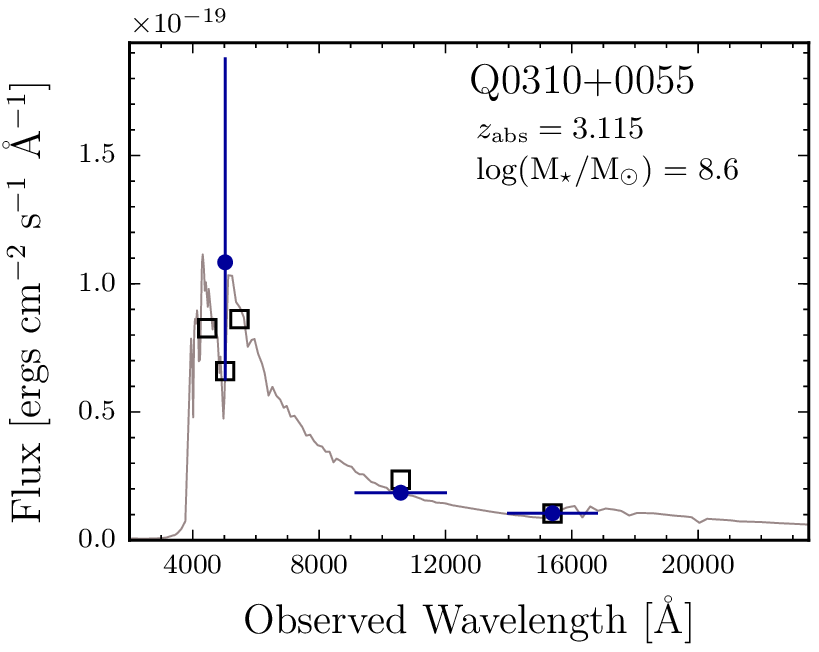}
    \includegraphics[width=0.38\textwidth]{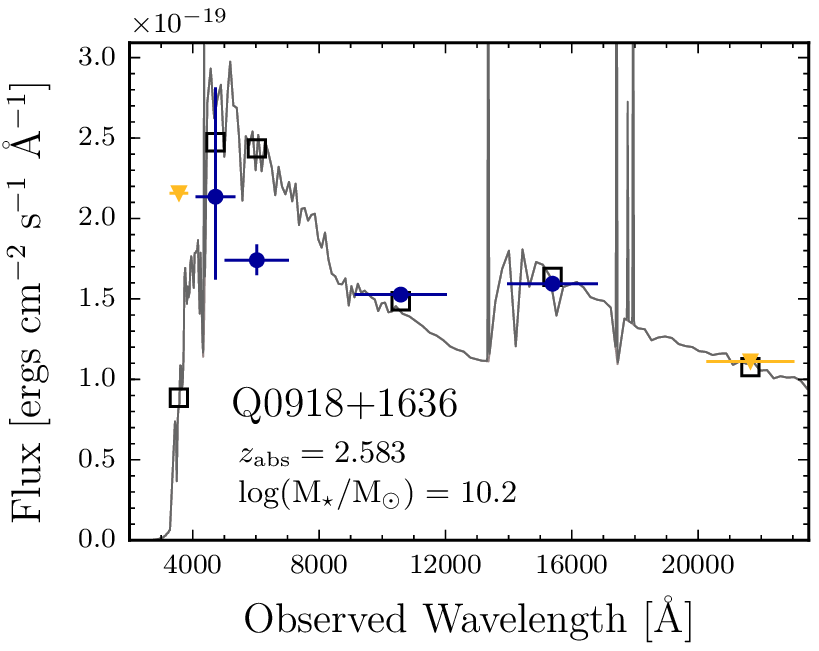}
    \includegraphics[width=0.38\textwidth]{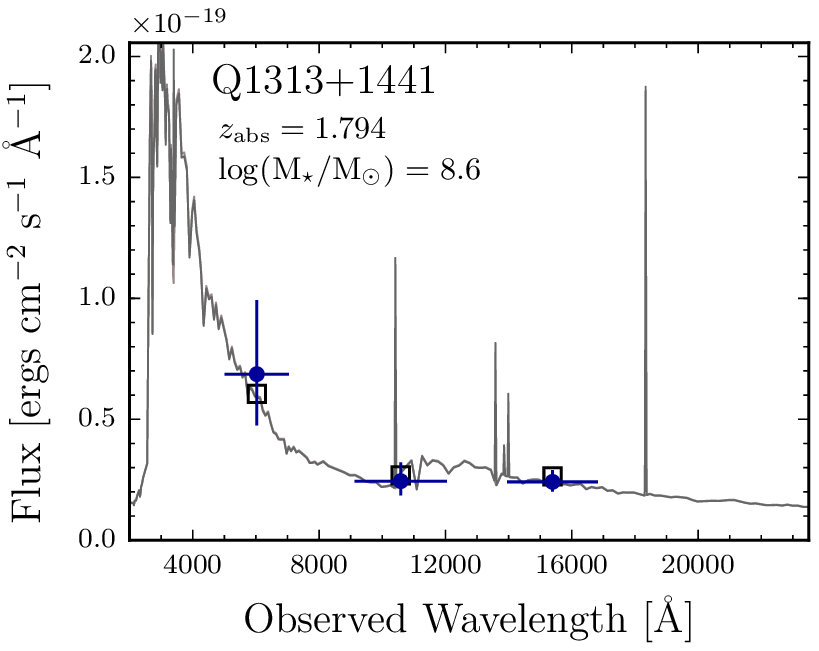}
    \includegraphics[width=0.38\textwidth]{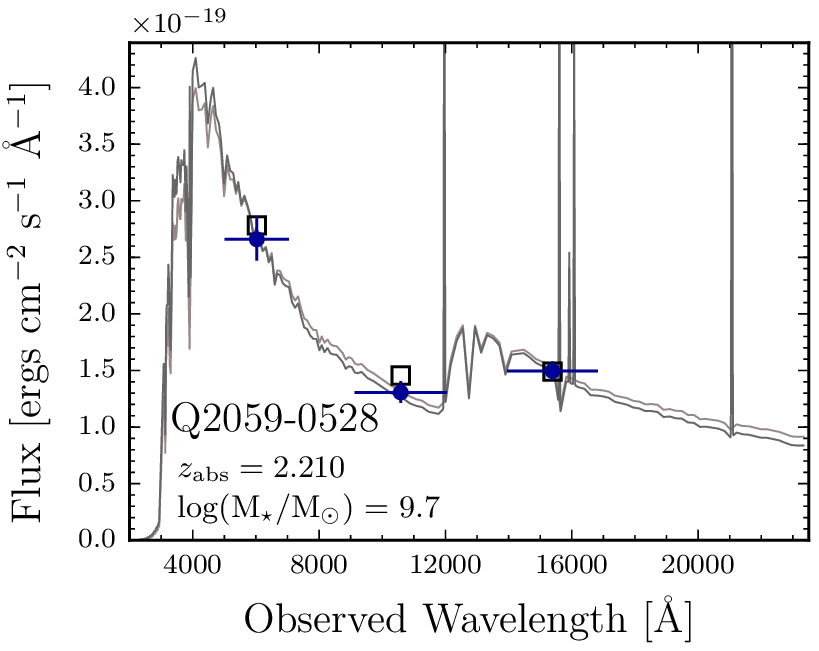}
    \includegraphics[width=0.38\textwidth]{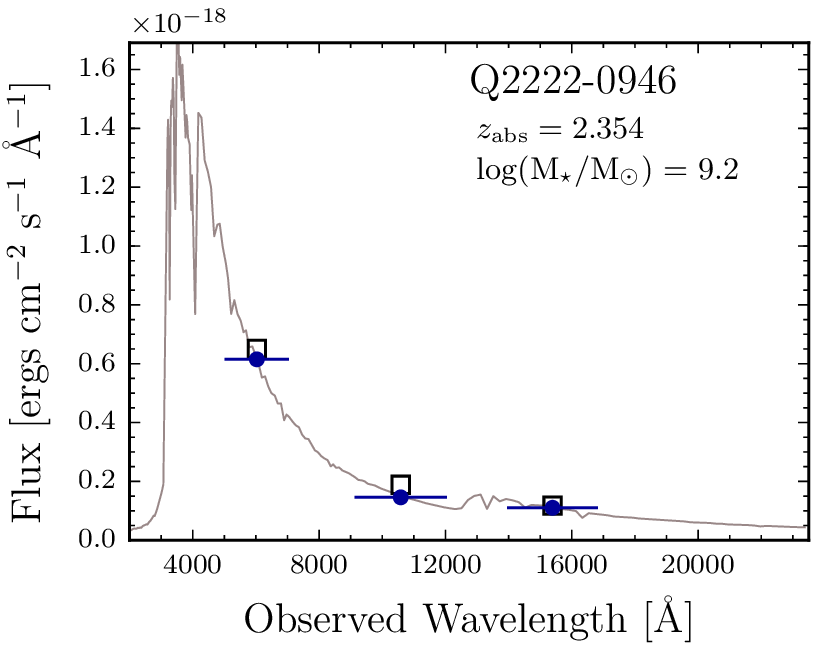}
    \includegraphics[width=0.38\textwidth]{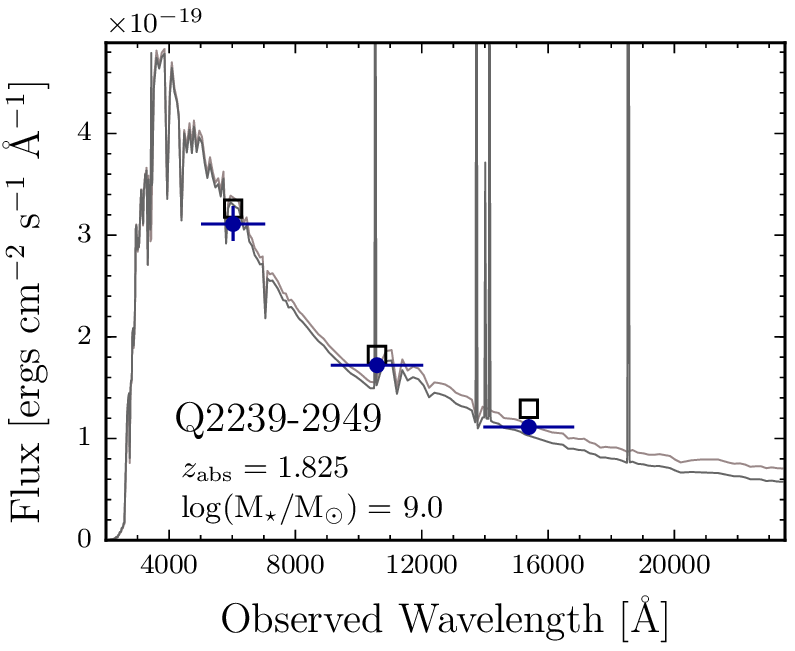}
    \includegraphics[width=0.38\textwidth]{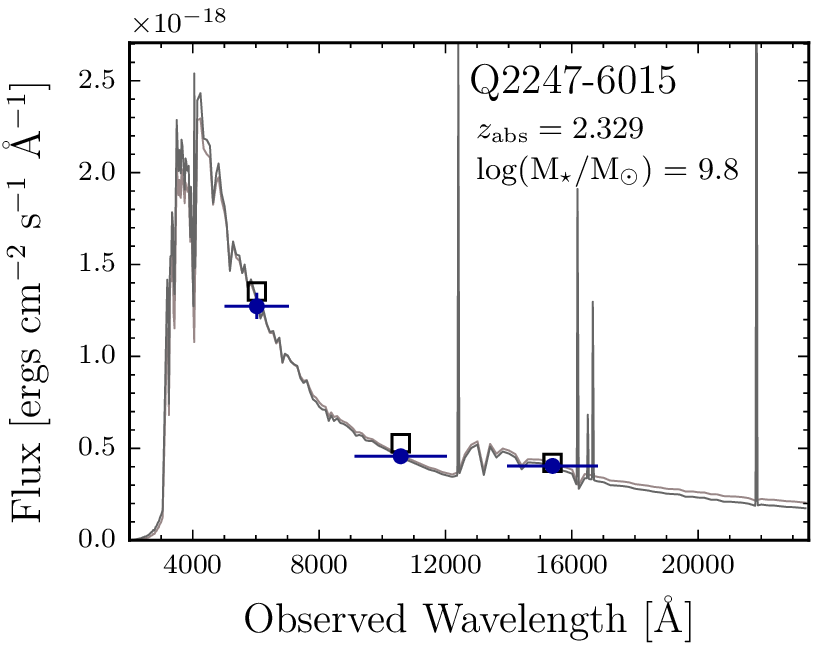}
\caption{Spectral energy distribution (SED) for the objects where a
  fit was possible. Blue (yellow) filled symbols refer to included
  measurements (limits), with horizontal error-bars indicating the
  FWHM of each filter. Grey empty squares show the best-fit
  transmission-weighted filter-flux. The two SEDs in each panel show
  the best fit with and without nebular emission. In many cases the
  two fits overlap, indicating the robustness of the solution. Each
  panel displays target name, absorber redshift, and the SED-based
  stellar mass of the counterpart. The conservative B- and V-band
  non-detections reported in \citet{Kashikawa2014} were included as
  upper limits during fitting, but lie above the Q0310$+$0055 panel's
  displayed flux range.}
\label{fig:SEDs}
\end{figure*}

\subsection{Modelling the spectral energy distribution}
\label{subsec:SED}

We determine the stellar mass (M$_{\star}$) for each of the absorbing
galaxies by fitting the SEDs using the code {\sc LePhare}
\citep{Arnouts1999, Ilbert2006}.  For each target, we fix the redshift
and apply the corrections for Galactic extinction to the magnitudes.
{\sc LePhare} fits the SEDs by minimising $\chi^2$ across a user
defined grid of parameters. As input, we use standard
\citet[][BC03]{Bruzual2003} single stellar population (SSP) templates
for a \cite{Chabrier2003} IMF, with default model metallicities
$Z/Z_{\odot} = $ 0.2, 0.4 and 1.0.  We assume exponentially declining
star formation histories with $e$-folding time-scales spanning 0.1 --
30 Gyrs and limit the stellar population ages to span $0 -
4~\mathrm{Gyrs}$, which corresponds to the age of the Universe at the
lowest absorption-redshift in our sample.  We adopt a
\cite{Calzetti2000} attenuation-curve as we are probing redshifts
around the peak of cosmic star formation. The colour excess,
$E_{B-V}$, is sampled in steps of 0.05 in the range $0 - 0.3$~mag. The
range is extended, as needed, to ensure that the preferred $E_{B-V}$
is associated with a $\chi^2$-minimum rather than a grid-boundary.
 
The resulting SED fits of the absorbing galaxies are shown in
Fig.~\ref{fig:SEDs} and the resulting colour excesses, SFRs and
stellar masses are reported in Table~\ref{tab:SFR}.

All values are based on SED fits including the nebular emission lines
apart from Q\,2222--0946 and Q\,0310+0055, for which information on
spectroscopic emission-line fluxes was used to isolate the stellar
continuum \citep[see][respectively]{Krogager2013,Kashikawa2014}.  For
completeness, we also overplot the best-fit SEDs using the same grid,
but excluding emission lines.

\subsubsection{Comparison to literature measurements}

In the case of Q\,2222--0946, we correct our F160W broadband magnitude
for a 33\% nebular emission-line contribution, as determined by
\citet{Krogager2013}.  This yields results which are in good agreement
with those presented by \citet{Krogager2013}, despite their use of a
randomised library of different star formation histories instead of a
parameterised approach as we have assumed in this work.  The
consistency between these results is contrasted by the measurement of
$\log(\mathrm{M}_{\star}/\mathrm{M}_{\odot}) = 9.7\pm 0.3$ reported by
\cite{Augustin2018}. Similar to our work, the authors use {\sc
  LePhare} for their analysis; However, our parameter-grids differ in
important aspects. In particular, their models use a single burst of
star formation whereas we allow for a range of star formation
histories.  It is also unclear (i) how they include dust reddening
(both in range and sampling); (ii) whether they correct their
magnitude measurements, adopted from \citet{Krogager2013}, for
Galactic extinction; and (iii) whether the template fit included
nebular emission (as supported by the SED fits presented in Figure 5
by \citet{Augustin2018}), or not (as supported by Table 3 of
\citet{Augustin2018} in which they report the F160W nebular-emission
corrected continuum magnitude of \citet{Krogager2013}).  Indeed, by
restricting the input models to a single burst population, using the
F160W magnitude which includes the emission-line flux reported by
\citet{Krogager2013}, and enabling {\sc LePhare}'s nebular emission
prescription, we are able to retrieve a stellar mass of
$\log(\mathrm{M}_{\star}/\mathrm{M}_{\odot}) = 9.6\pm 0.3$ which is in
closer agreement with the value reported by \citet{Augustin2018},
whilst maintaining a near identical star formation rate of $\log
(\text{SFR} / \mathrm{M}_{\odot} ~ \mathrm{yr}^{-1}) = 0.9\pm 0.2$ to
that which we report based on our own grid (see Table \ref{tab:SFR}).

Our magnitude measurements reported for Q\,0918+1636 (see
Table~\ref{tab:galfit}) appear in tension with those reported by
\citet[][their table 2]{Fynbo2013}.  Whereas our values refer to the
measured observables, \citet[][]{Fynbo2013} tabulated the magnitudes
after applying Galactic extinction corrections.  Once these
corrections are applied to our measurements
(Sect.~\ref{subsec:archivaldata}), all magnitudes are consistent
within $\lesssim 1\sigma$.  Reassuringly, the stellar mass of the DLA
counterpart derived in this work is in perfect agreement with the
value of $\log(\mathrm{M}_{\star} /
\mathrm{M}_{\odot})=10.1^{+0.2}_{-0.1}$ reported by \citet[][their
  Table 3]{Fynbo2013}.

\begin{figure*}
    \centering
    \includegraphics[width=0.48\textwidth]{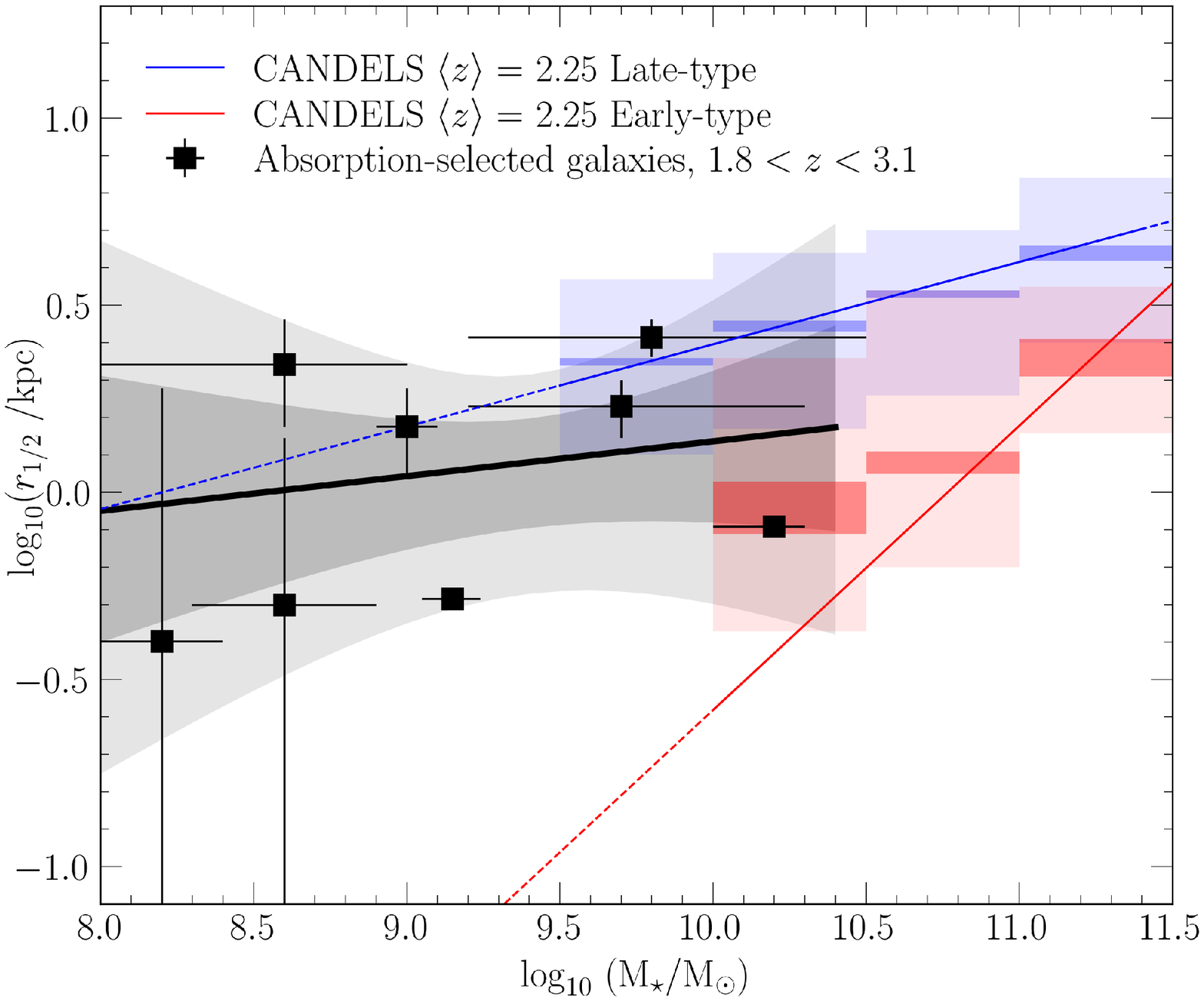}  
    \includegraphics[width=0.485\textwidth]{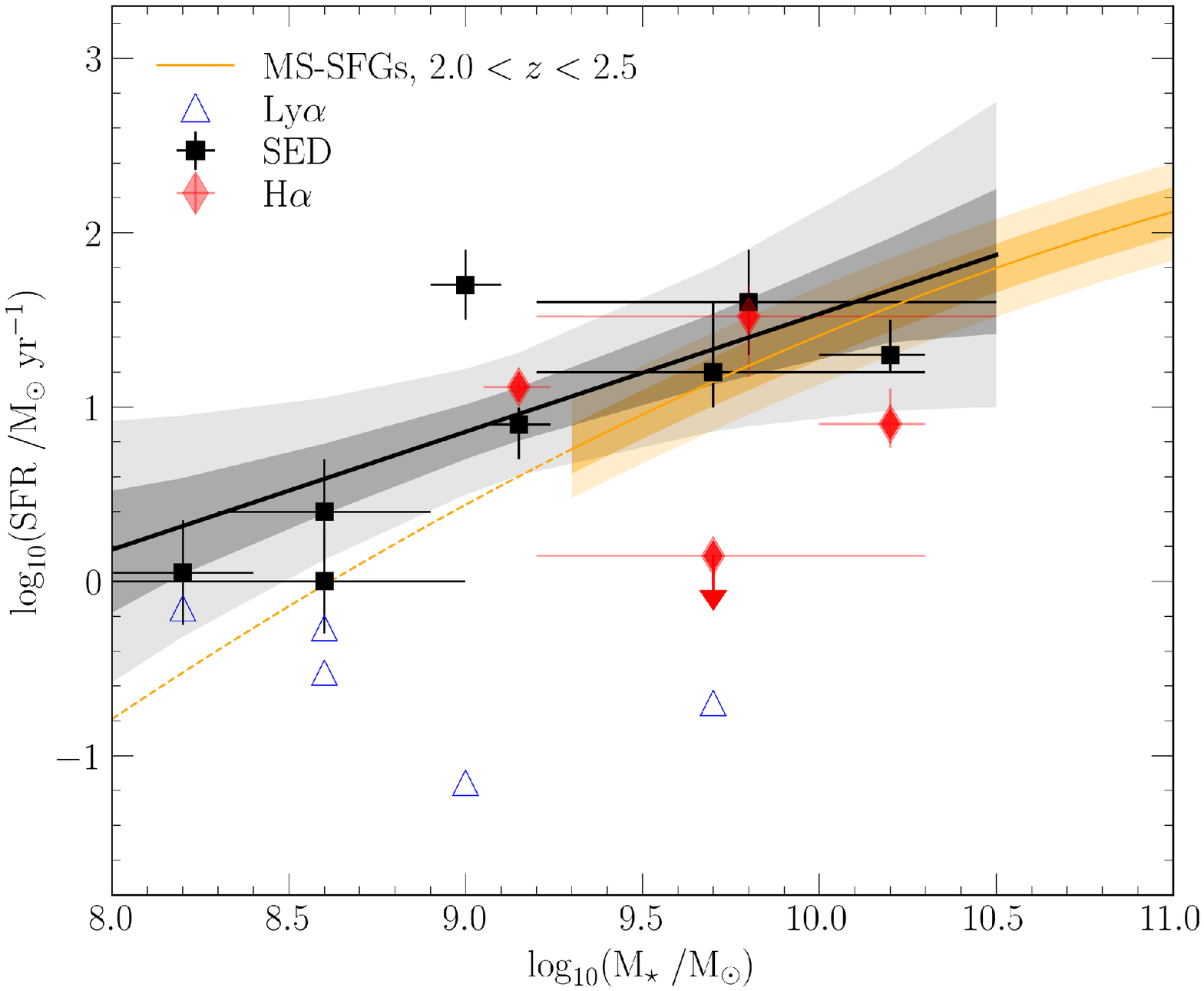}  
\caption{The mass--size relation (left) and the `main sequence' of
  star-forming galaxies (right) of our sample of absorption-selected
  galaxies at $z\sim 2.3$ observed with \HST. The absorption-selected
  galaxies preferentially select the star-forming galaxy population,
  with an extension to lower stellar masses. In both panels, the black
  line and its associated grey shaded regions depict the best fit
  relation and the $1\sigma -$ and $2\sigma$ confidence intervals to
  the absorption-selected galaxies presented in this work. The fit in
  the right panel uses only the SED-based SFRs (black squares).  In
  the left panel, we show the scaling relations observed for
  luminosity-selected samples of early-type (red) and late-type (blue)
  galaxies at redshifts $\langle z \rangle = 2.25$ by
  \citet{vanderWel2014}, with dark shaded regions representing medians
  and their errors and the light shaded regions representing the 16-
  and 84-percentile ranges.  Solid lines show the relations derived
  from effective semi-major axis measurements, with dashed segments
  illustrating an extrapolation beyond the reported completeness
  limits.  The box-shaped shaded regions show the mean and $1\sigma$
  scatter of circularised radii, whose definition is closer to our
  half-light radii. In the right panel, we show the `main sequence' of
  star-forming galaxies at redshifts $2.0<z<2.5$ (MS-SFGs, yellow
  line) from \citet{Whitaker2014}, with yellow shaded regions
  indicating the $1\sigma$ and $2\sigma$ confidence intervals
  including the scatter in the data from \citet{Whitaker2015}.  The
  extrapolation to lower masses is illustrated by the dashed segments.
  The black squares show SFR measurements based on the SED fit whereas
  red diamonds show spectroscopic measurements based on H$\alpha$. The
  blue triangles represent SFR lower limits based on \lya.  }
\label{fig:scalingrelations}
\end{figure*}

\section{Discussion}
\label{sec:discussion}

Having analysed the emission properties of our sample of high-redshift
galaxies associated with strong \HI\ absorbers, we can now investigate
how these objects compare to samples of luminosity-selected
galaxies. In Fig.~\ref{fig:scalingrelations} we show the established
mass--size relation \citep{vanderWel2014} and the so-called `main
sequence' of star-forming galaxies (MS-SFGs) from
\citet{Whitaker2014}, both constructed from luminosity-selected
samples at similar redshifts to our sample, which has a mean redshift
$\langle z_{\rm abs}\rangle = 2.3$. A more detailed comparison of our
\HST\ sample to each of these relations is presented below.

\subsection{Mass--size relation}
\citet{vanderWel2014} investigate the stellar-mass--size relation for
early- and late-type galaxies over a large range in redshift based on
the 3D-\emph{HST}/CANDELS survey. Early- and late-type galaxies are
separated by colour-criteria. The authors derive stellar masses
assuming the same initial mass function \citep{Chabrier2003} as we do
in this work.  \citeauthor{vanderWel2014} report their size estimates
in two ways; as the effective semi-major axis ($a_e$) from S\'ersic
models; and as circularised effective radii. In
Fig.~\ref{fig:scalingrelations}, we indicate the mass--$a_e$ relations
for early- and late-type galaxies at $\langle z \rangle=2.25$ reported
by \citet{vanderWel2014} as the red and blue lines,
respectively. Below the reported completeness limits of their work, we
show an extrapolation of these relations as dashed lines.  The
circularised effective radii by \citeauthor{vanderWel2014} are shown
as shaded boxes, where the vertical extent depicts the $1\sigma$
dispersion within a given stellar mass bin, and the horizontal extent
depicts the stellar mass range of the bin. We note that the
circularised radii, in definition, more closely resemble our
non-parametric approach, and should therefore make a more fair
comparison to the $r_{1/2}$ size-estimates of the absorption-selected
galaxy sample presented in this work.

In Fig.~\ref{fig:scalingrelations}, we also show our sample of
absorption-selected galaxies. We fit a mass--size relation to our
sample with a power-law, similar to \citealt{vanderWel2014}. For the
fit, we adopt the minimisation method described in \citet{Moller2013}
including a term for the intrinsic scatter, and taking into account
the asymmetric uncertainties in $\log({\rm M}_{\star} /
\mathrm{M}_{\odot})$ and measured radii. We obtain the following best
fit relation:

$$\log(r_{\mathrm{1/2}}/{\rm kpc}) = (-0.8_{-2.0}^{+2.3}) + (0.1\pm0.2) \log({\rm M}_{\star}/{\rm M_{\odot}})~,$$

\noindent with an internal scatter of $\sigma_{\log r_{1/2}} =
0.26$~dex.  The best-fit relation for our sample is shown as the solid
black line with $1\sigma$- and $2\sigma$ confidence intervals shown as
grey shaded regions around the line.  The relation inferred from our
sample matches more closely the relation for late-type (blue) galaxies
than the one for early-type (red) galaxies.  Quantitatively, we assess
the similarity between our sample and the two relations for early- and
late-type galaxies using the reduced $\chi^2$ statistic. For eight
degrees of freedom ($\nu=8$) and assuming a simple extrapolation of
the relations from luminosity-selected galaxies, we obtain $\chi^2 /
\nu = 9.5/8 = 1.2$ to the blue line, and $\chi^2 / \nu = 70/8 = 8.8$
to the red line.  Given the number of degrees of freedom, the
expectation value of the reduced $\chi^2$ statistic is $\chi^2/\nu
\approx 1 \pm \sqrt{2/\nu} \approx 1.0 \pm 0.5$.  We therefore
conclude that our sample is consistent (to within $<1\sigma$) with the
relation for late-type galaxies and inconsistent with the relation for
early-type galaxies at more than $15\sigma$.

Albeit limited by sample size and individual measurement
uncertainties, it is encouraging to see that, even towards the
low-mass end, our absorption-selected sample follows the extrapolated
luminosity-selected mass-size relation for late-type galaxies at
$z\sim2$. This result is consistent with the notion that identifying
galaxies in absorption preferentially selects faint, gas-rich,
star-forming galaxies \citep[e.g.,][]{Fynbo1999}. Remarkably, more
than $50\%$ of our sample has stellar masses below the formal
completeness limit of the luminosity-selected \HST\ survey, despite
the fact that our metallicity cut pre-selects the most massive
(sub-)DLA galaxies known.

\subsection{Main sequence of star-forming galaxies}

In this section we consider only the sub-sample of 8 galaxies for
which we derive SED-based SFRs and stellar masses, as listed in Table
\ref{tab:SFR}.

In Fig.~\ref{fig:scalingrelations}, we show the polynomial
parameterization of the MS-SFGs for the luminosity-selected galaxies at
$z\sim 2-2.5$ reported by \citet{Whitaker2014}, with the associated
$1\sigma-$ and $2\sigma$ scatter in the relation based on the
intrinsic dispersion $\sigma_{\rm intrinsic} = 0.14$ reported in
\citet{Whitaker2015}.  For our sample, we have incomplete data from
three different SFR tracers: lower limits from \lya;
recombination-line measurements of H$\alpha$ which trace the
near-instantaneous SFR on time-scales of $\sim 10$~Myr; and SED-based
values (see Section \ref{subsec:SED}) that trace the ongoing SFR (see
the documentation of {\sc LePhare}).  To form a complete census, we
plot all the SFR tracers for each object in
Figure~\ref{fig:scalingrelations}.

The results of the various diagnostics can be summarised as
follows. For objects with SFR measurements in \lya\ and either
H$\alpha$ or SED, the lower limits from \lya\ are consistent with the
other tracers in all cases.  Considering the limited number of
photometric bands available to constrain our SED fits, we find a good
agreement between H$\alpha$ and SED based SFRs where both are
available, with the notable exception of Q\,2059--0528. For this
object, expressed in terms of logarithmic SFR, our SED based
measurement is in 2.6$\sigma$ tension with the H$\alpha$ upper limit
reported by \citet{Peroux2012}. We note, however, that the H$\alpha$
limit may be underestimated due to the added uncertainties introduced
by the quasar PSF and its subtraction. In addition, the characteristic
SFR timescales differ, and our \emph{HST} filter selection is
optimised for the determination of SED-based stellar masses, not of
SED-based SFRs. We therefore resort to the SED-based SFR on a
statistical level of the sample, but emphasise caution not to
overinterpret the SED-based SFR for individual objects.

Since the SED-based SFRs are available for all galaxies in the
sub-sample treated in this section, we use these SFR estimates to
investigate the relation between stellar mass and star formation. A
log-linear (i.e., power-law) fit of the stellar-mass--SFR$_{\rm SED}$
relation using an orthogonal linear regression method with asymmetric
uncertainties to our sample yields:
$$\log({\rm SFR}/ \MsunYr) = (0.7 \pm 0.5)\log({\rm M_{\star}/M_{\odot}}) - (5.2 \pm 4.3)~.$$

In Fig. \ref{fig:scalingrelations} we plot the best-fit relation to
our sample as a solid black line, and its $1\sigma$ and $2\sigma$
confidence intervals as grey shaded regions around the line.

The MS-SFG relation at $2.0<z<2.5$ by \cite{Whitaker2014} suggests a
linear relation with a steepening slope of order unity towards low
stellar masses, consistent to $<1\sigma$ with our fitted
relation. Whereas this comparison is based on an extrapolation of the
\citet{Whitaker2014} relation below the reported mass completeness
limit, our results are furthermore consistent with
\citet{Kochiashvili2015}, who report a log-linear slope of $1.17$
based on an emission-line selected galaxy sample at $z \sim 1.85$ with
measurements probing low stellar masses in the range of $8.5 <
\log({\rm M_{\star}/M_{\odot}}) < 9.4$.

The fact that our sample matches the established MS-SFGs is
interesting, when comparing to lower redshifts, where
absorption-selected galaxies are found to lie \emph{below} the `main
sequence' at $z\sim 0.7$ \citep{Moller2018,Kanekar2018,Rhodin2018}.
The differences hint at a redshift evolution in the way
absorption-selection traces the underlying galaxy population.
However, our high-redshift \emph{HST} sample probes lower stellar
masses on average
($\langle \log {\rm M_{\star} / M_{\odot}} \rangle = 9.2$)
than the low-redshift sample
($\langle \log {\rm M_{\star} / M_{\odot}} \rangle = 10.0$;
\citealt{Rhodin2018}). This inhibits us from discriminating any
evolution with redshift from an evolution with stellar mass.

\section{Conclusions} \label{sec:conclusions}
In this work, we performed a systematic analysis of high-redshift ($2
\lesssim z \lesssim 3$) galaxies associated with strong
\HI\ ($\logNHIcm > 19$) absorbers, for which the emission counterparts
were known in advance based on spectroscopic emission line
identifications. In seven fields, we obtained new \emph{HST/WFC3}
imaging data. In addition, we re-analysed three fields with archival
\HST\ images of similar configuration and quality, which renders a
total homogeneous sample of ten fields compiled from two \emph{HST}
campaigns.

The high spatial resolution of the \HST\ images combined with a
careful subtraction of the quasar PSFs allow us to robustly detect
continuum emission counterparts for nine systems in at least one
filter (seven new detections; two confirmations of previously
published detections), and to report one tentative detection in the
Q0338--0005 field.

Accounting for the quasar PSFs, each absorbing galaxy is modelled with
a multi-component S\'ersic model to describe its light
distribution. This enables us to measure broad-band magnitudes and
half-light radii. Combined with redshift, known from spectroscopic
observations in absorption and emission, the detection of the galaxy
in more than one filter allows us to derive SED-based stellar masses
and star formation rates for eight out of the ten targets. The main
results can be summarised as follows:

\begin{itemize}

    \item With stellar masses of $\log ({\rm M_{\star} / M}_{\odot}) =
      8 - 10$ and half-light radii of $r_{1/2} = 0.4 - 2.6$~kpc, our
      sample forms a relation consistent with the mass--size relation
      for luminosity-selected, late-type galaxies at $z=2.25$
      \citep{vanderWel2014}, extrapolated towards the faint end of the
      galaxy luminosity function by more than one order of magnitude
      in stellar mass.

    \item Combining the stellar masses with SFR estimates of $1 -
      50$~\MsunYr\ based on spectroscopic H$\alpha$ emission line
      fluxes and SED-fits, our sample forms a relation consistent with
      the MS-SFGs derived for luminosity-selected samples at similar
      redshifts \citep{Whitaker2014}, extrapolated to ten times lower
      stellar masses.

    \item With $\sim 40\%$ of our sample displaying complex light
      distributions, whose modelling requires multiple S\'ersic
      components, absorption-selected galaxies at $z \sim 2 -3$ are
      consistent with the findings of clumpy morphologies in
      high-redshift galaxies \citep{Livermore2015}.
\end{itemize}

Lastly, we revisit the absorption-line metallicity of the DLA towards
Q\,0139--0824 using VLT/X-Shooter data. We furthermore provide new
measurements of the spectroscopic detections of emission counterparts
of the absorbers towards Q\,0139--0824 and Q\,0124+0044 using archival
VLT/X-Shooter and VLT/FORS1 data.

Based on our sample of absorption-selected, high-redshift galaxies we
suggest that, at redshift $2-3$, galaxies associated to strong
\HI\ absorbers predominantly trace gas-rich, late-type, star-forming
galaxies from the faint end of the Lyman-break galaxy luminosity
function.  Previous analyses of a smaller sample led to similar
conclusions \citep{Moller2002}. With the \HST\ observations presented
in this work, we demonstrate that such galaxies follow scaling
relations between stellar mass, SFR, and size established for
luminosity-selected samples, and extend these relations to lower
masses by one order of magnitude.

\section*{acknowledgments}
    NHPR and LC are supported by the Independent Research Fund Denmark
    (DFF - 4090--00079).  JKK acknowledges support from the Danish
    Council for Independent Research (EU-FP7 under Marie-Curie grant
    agreement no. 600207; DFF-MOBILEX -- 5051-00115).  KEH
    acknowledges support by a Project Grant (162948--051) from The
    Icelandic Research Fund. FV acknowledges support from the
    Carlsberg Foundation Research Grant CF18-0388 ``Galaxies: Rise and
    Death'' and the Cosmic Dawn Center of Excellence funded by the
    Danish National Research Foundation under the Grant No. 140.

\section*{Data availability}
    Based on observations made with the NASA/ESA \emph{Hubble Space
      Telescope}, obtained at the Space Telescope Science Institute,
    which is operated by the Association of Universities for Research
    in Astronomy, Inc., under NASA contract NAS 5-26555.  Data
    obtained for this article were accessed via
    https://archive.stsci.edu/hst/ (Programme ID 12553).

    Based on archival FORS1 and X-Shooter observations collected at
    the European Southern Observatory under ESO programme IDs
    189.A-0424 and 081.A-0506, accessed via http://archive.eso.org.

    This research has made use of the SIMBAD database
    (\url{http://simbad.u-strasbg.fr/simbad/}), operated at CDS, Strasbourg,
    France \citep{Simbad2000}.  Astropy \citep{Astropy2013}, Photutils
    \citep{Photutils2016}, Matplotlib \citep{Matplotlib2007}, and
    Drizzlepac \citep{Drizzlepac2012}.

%


\bibliographystyle{mnras}
\bibliography{HST2020.bib}

\begin{thebibliography}{}
\makeatletter
\relax
\def\mn@urlcharsother{\let\do\@makeother \do\$\do\&\do\#\do\^\do\_\do\%\do\~}
\def\mn@doi{\begingroup\mn@urlcharsother \@ifnextchar [ {\mn@doi@}
  {\mn@doi@[]}}
\def\mn@doi@[#1]#2{\def\@tempa{#1}\ifx\@tempa\@empty \href
  {http://dx.doi.org/#2} {doi:#2}\else \href {http://dx.doi.org/#2} {#1}\fi
  \endgroup}
\def\mn@eprint#1#2{\mn@eprint@#1:#2::\@nil}
\def\mn@eprint@arXiv#1{\href {http://arxiv.org/abs/#1} {{\tt arXiv:#1}}}
\def\mn@eprint@dblp#1{\href {http://dblp.uni-trier.de/rec/bibtex/#1.xml}
  {dblp:#1}}
\def\mn@eprint@#1:#2:#3:#4\@nil{\def\@tempa {#1}\def\@tempb {#2}\def\@tempc
  {#3}\ifx \@tempc \@empty \let \@tempc \@tempb \let \@tempb \@tempa \fi \ifx
  \@tempb \@empty \def\@tempb {arXiv}\fi \@ifundefined
  {mn@eprint@\@tempb}{\@tempb:\@tempc}{\expandafter \expandafter \csname
  mn@eprint@\@tempb\endcsname \expandafter{\@tempc}}}

\bibitem[\protect\citeauthoryear{{Abbott} et~al.,}{{Abbott}
  et~al.}{2000}]{Abbott2000}
{Abbott} T.~M.,  et~al., 2000, in {Iye} M.,  {Moorwood} A.~F.,  eds,  \procspie
  Vol. 4008, Optical and IR Telescope Instrumentation and Detectors. pp
  714--719, \mn@doi{10.1117/12.395528}

\bibitem[\protect\citeauthoryear{{Arnouts}, {Cristiani}, {Moscardini},
  {Matarrese}, {Lucchin}, {Fontana}  \& {Giallongo}}{{Arnouts}
  et~al.}{1999}]{Arnouts1999}
{Arnouts} S.,  {Cristiani} S.,  {Moscardini} L.,  {Matarrese} S.,  {Lucchin}
  F.,  {Fontana} A.,   {Giallongo} E.,  1999, \mn@doi [\mnras]
  {10.1046/j.1365-8711.1999.02978.x}, \href
  {http://adsabs.harvard.edu/abs/1999MNRAS.310..540A} {310, 540}

\bibitem[\protect\citeauthoryear{{Asplund}, {Grevesse}, {Sauval}  \&
  {Scott}}{{Asplund} et~al.}{2009}]{Asplund2009}
{Asplund} M.,  {Grevesse} N.,  {Sauval} A.~J.,   {Scott} P.,  2009, \mn@doi
  [\araa] {10.1146/annurev.astro.46.060407.145222}, \href
  {https://ui.adsabs.harvard.edu/abs/2009ARA&A..47..481A} {47, 481}

\bibitem[\protect\citeauthoryear{{Astropy Collaboration} et~al.,}{{Astropy
  Collaboration} et~al.}{2013}]{Astropy2013}
{Astropy Collaboration} et~al., 2013, \mn@doi [\aap]
  {10.1051/0004-6361/201322068}, \href
  {http://adsabs.harvard.edu/abs/2013A%26A...558A..33A} {558, A33}

\bibitem[\protect\citeauthoryear{{Augustin} et~al.,}{{Augustin}
  et~al.}{2018}]{Augustin2018}
{Augustin} R.,  et~al., 2018, \mn@doi [\mnras] {10.1093/mnras/sty1287}, \href
  {https://ui.adsabs.harvard.edu/abs/2018MNRAS.478.3120A} {478, 3120}

\bibitem[\protect\citeauthoryear{{Bashir}, {Zafar}, {Khan}  \&
  {Chishtie}}{{Bashir} et~al.}{2019}]{Bashir2019}
{Bashir} W.,  {Zafar} T.,  {Khan} F.~M.,   {Chishtie} F.,  2019, \mn@doi [New
  Astronomy] {10.1016/j.newast.2018.07.001}, \href
  {https://ui.adsabs.harvard.edu/abs/2019NewA...66....9B} {66, 9}

\bibitem[\protect\citeauthoryear{{Berg} et~al.,}{{Berg}
  et~al.}{2016}]{Berg2016}
{Berg} T.~A.~M.,  et~al., 2016, \mn@doi [\mnras] {10.1093/mnras/stw2232}, \href
  {http://adsabs.harvard.edu/abs/2016MNRAS.463.3021B} {463, 3021}

\bibitem[\protect\citeauthoryear{{Berry}, {Somerville}, {Gawiser}, {Maller},
  {Popping}  \& {Trager}}{{Berry} et~al.}{2016}]{Berry2016}
{Berry} M.,  {Somerville} R.~S.,  {Gawiser} E.,  {Maller} A.~H.,  {Popping} G.,
    {Trager} S.~C.,  2016, \mn@doi [\mnras] {10.1093/mnras/stw231}, \href
  {http://adsabs.harvard.edu/abs/2016MNRAS.458..531B} {458, 531}

\bibitem[\protect\citeauthoryear{{Bouch{\'e}} et~al.,}{{Bouch{\'e}}
  et~al.}{2012}]{Bouche2012}
{Bouch{\'e}} N.,  et~al., 2012, \mn@doi [\mnras]
  {10.1111/j.1365-2966.2011.19500.x}, \href
  {http://adsabs.harvard.edu/abs/2012MNRAS.419....2B} {419, 2}

\bibitem[\protect\citeauthoryear{{Bouch{\'e}}, {Murphy}, {Kacprzak},
  {P{\'e}roux}, {Contini}, {Martin}  \& {Dessauges-Zavadsky}}{{Bouch{\'e}}
  et~al.}{2013}]{Bouche2013}
{Bouch{\'e}} N.,  {Murphy} M.~T.,  {Kacprzak} G.~G.,  {P{\'e}roux} C.,
  {Contini} T.,  {Martin} C.~L.,   {Dessauges-Zavadsky} M.,  2013, \mn@doi
  [Science] {10.1126/science.1234209}, \href
  {http://adsabs.harvard.edu/abs/2013Sci...341...50B} {341, 50}

\bibitem[\protect\citeauthoryear{{Bradley} et~al.,}{{Bradley}
  et~al.}{2016}]{Photutils2016}
{Bradley} L.,  et~al., 2016, {Photutils: Photometry tools}, Astrophysics Source
  Code Library (\mn@eprint {ascl} {1609.011})

\bibitem[\protect\citeauthoryear{{Bruzual} \& {Charlot}}{{Bruzual} \&
  {Charlot}}{2003}]{Bruzual2003}
{Bruzual} G.,  {Charlot} S.,  2003, \mn@doi [\mnras]
  {10.1046/j.1365-8711.2003.06897.x}, \href
  {http://adsabs.harvard.edu/abs/2003MNRAS.344.1000B} {344, 1000}

\bibitem[\protect\citeauthoryear{{Calzetti}, {Armus}, {Bohlin}, {Kinney},
  {Koornneef}  \& {Storchi-Bergmann}}{{Calzetti} et~al.}{2000}]{Calzetti2000}
{Calzetti} D.,  {Armus} L.,  {Bohlin} R.~C.,  {Kinney} A.~L.,  {Koornneef} J.,
   {Storchi-Bergmann} T.,  2000, \mn@doi [\apj] {10.1086/308692}, \href
  {http://adsabs.harvard.edu/abs/2000ApJ...533..682C} {533, 682}

\bibitem[\protect\citeauthoryear{{Cappetta}, {D'Odorico}, {Cristiani}, {Saitta}
   \& {Viel}}{{Cappetta} et~al.}{2010}]{Cappetta2010}
{Cappetta} M.,  {D'Odorico} V.,  {Cristiani} S.,  {Saitta} F.,   {Viel} M.,
  2010, \mn@doi [\mnras] {10.1111/j.1365-2966.2010.16981.x}, \href
  {http://adsabs.harvard.edu/abs/2010MNRAS.407.1290C} {407, 1290}

\bibitem[\protect\citeauthoryear{{Cashman}, {Kulkarni}, {Kisielius}, {Ferland}
  \& {Bogdanovich}}{{Cashman} et~al.}{2017}]{Cashman2017}
{Cashman} F.~H.,  {Kulkarni} V.~P.,  {Kisielius} R.,  {Ferland} G.~J.,
  {Bogdanovich} P.,  2017, \mn@doi [\apjs] {10.3847/1538-4365/aa6d84}, \href
  {http://adsabs.harvard.edu/abs/2017ApJS..230....8C} {230, 8}

\bibitem[\protect\citeauthoryear{{Chabrier}}{{Chabrier}}{2003}]{Chabrier2003}
{Chabrier} G.,  2003, \mn@doi [\pasp] {10.1086/376392}, \href
  {http://adsabs.harvard.edu/abs/2003PASP..115..763C} {115, 763}

\bibitem[\protect\citeauthoryear{{Christensen}, {M{\o}ller}, {Fynbo}  \&
  {Zafar}}{{Christensen} et~al.}{2014}]{Christensen2014}
{Christensen} L.,  {M{\o}ller} P.,  {Fynbo} J.~P.~U.,   {Zafar} T.,  2014,
  \mn@doi [\mnras] {10.1093/mnras/stu1726}, \href
  {http://adsabs.harvard.edu/abs/2014MNRAS.445..225C} {445, 225}

\bibitem[\protect\citeauthoryear{{Crighton} et~al.,}{{Crighton}
  et~al.}{2015}]{Crighton2015}
{Crighton} N.~H.~M.,  et~al., 2015, \mn@doi [\mnras] {10.1093/mnras/stv1182},
  \href {http://adsabs.harvard.edu/abs/2015MNRAS.452..217C} {452, 217}

\bibitem[\protect\citeauthoryear{{De Cia}, {Ledoux}, {Mattsson}, {Petitjean},
  {Srianand}, {Gavignaud}  \& {Jenkins}}{{De Cia} et~al.}{2016}]{DeCia2016}
{De Cia} A.,  {Ledoux} C.,  {Mattsson} L.,  {Petitjean} P.,  {Srianand} R.,
  {Gavignaud} I.,   {Jenkins} E.~B.,  2016, \mn@doi [\aap]
  {10.1051/0004-6361/201527895}, \href
  {http://adsabs.harvard.edu/abs/2016A%26A...596A..97D} {596, A97}

\bibitem[\protect\citeauthoryear{{Fynbo}, {M{\o}ller}  \& {Warren}}{{Fynbo}
  et~al.}{1999}]{Fynbo1999}
{Fynbo} J.~U.,  {M{\o}ller} P.,   {Warren} S.~J.,  1999, \mnras, \href
  {http://adsabs.harvard.edu/abs/1999MNRAS.305..849F} {305, 849}

\bibitem[\protect\citeauthoryear{{Fynbo}, {Prochaska}, {Sommer-Larsen},
  {Dessauges-Zavadsky}  \& {M{\o}ller}}{{Fynbo} et~al.}{2008}]{Fynbo2008}
{Fynbo} J.~P.~U.,  {Prochaska} J.~X.,  {Sommer-Larsen} J.,
  {Dessauges-Zavadsky} M.,   {M{\o}ller} P.,  2008, \mn@doi [\apj]
  {10.1086/589555}, \href {http://adsabs.harvard.edu/abs/2008ApJ...683..321F}
  {683, 321}

\bibitem[\protect\citeauthoryear{{Fynbo} et~al.,}{{Fynbo}
  et~al.}{2010}]{Fynbo2010}
{Fynbo} J.~P.~U.,  et~al., 2010, \mn@doi [\mnras]
  {10.1111/j.1365-2966.2010.17294.x}, \href
  {http://adsabs.harvard.edu/abs/2010MNRAS.408.2128F} {408, 2128}

\bibitem[\protect\citeauthoryear{{Fynbo} et~al.,}{{Fynbo}
  et~al.}{2011}]{Fynbo2011}
{Fynbo} J.~P.~U.,  et~al., 2011, \mn@doi [\mnras]
  {10.1111/j.1365-2966.2011.18318.x}, \href
  {http://adsabs.harvard.edu/abs/2011MNRAS.413.2481F} {413, 2481}

\bibitem[\protect\citeauthoryear{{Fynbo} et~al.,}{{Fynbo}
  et~al.}{2013}]{Fynbo2013}
{Fynbo} J.~P.~U.,  et~al., 2013, \mn@doi [\mnras] {10.1093/mnras/stt1579},
  \href {http://adsabs.harvard.edu/abs/2013MNRAS.436..361F} {436, 361}

\bibitem[\protect\citeauthoryear{{Hartoog}, {Fynbo}, {Kaper}, {De Cia}  \&
  {Bagdonaite}}{{Hartoog} et~al.}{2015}]{Hartoog2015}
{Hartoog} O.~E.,  {Fynbo} J.~P.~U.,  {Kaper} L.,  {De Cia} A.,   {Bagdonaite}
  J.,  2015, \mn@doi [\mnras] {10.1093/mnras/stu2578}, \href
  {http://adsabs.harvard.edu/abs/2015MNRAS.447.2738H} {447, 2738}

\bibitem[\protect\citeauthoryear{{Hayes} et~al.,}{{Hayes}
  et~al.}{2010}]{Hayes2010}
{Hayes} M.,  et~al., 2010, \mn@doi [\nat] {10.1038/nature08881}, \href
  {http://adsabs.harvard.edu/abs/2010Natur.464..562H} {464, 562}

\bibitem[\protect\citeauthoryear{Hunter}{Hunter}{2007}]{Matplotlib2007}
Hunter J.~D.,  2007, \mn@doi [Computing In Science \& Engineering]
  {10.1109/MCSE.2007.55}, 9, 90

\bibitem[\protect\citeauthoryear{{Ilbert} et~al.,}{{Ilbert}
  et~al.}{2006}]{Ilbert2006}
{Ilbert} O.,  et~al., 2006, \mn@doi [\aap] {10.1051/0004-6361:20065138}, \href
  {http://adsabs.harvard.edu/abs/2006A%26A...457..841I} {457, 841}

\bibitem[\protect\citeauthoryear{{Jorgenson} \& {Wolfe}}{{Jorgenson} \&
  {Wolfe}}{2014}]{Jorgenson2014}
{Jorgenson} R.~A.,  {Wolfe} A.~M.,  2014, \mn@doi [\apj]
  {10.1088/0004-637X/785/1/16}, \href
  {https://ui.adsabs.harvard.edu/abs/2014ApJ...785...16J} {785, 16}

\bibitem[\protect\citeauthoryear{{Kanekar} et~al.,}{{Kanekar}
  et~al.}{2018}]{Kanekar2018}
{Kanekar} N.,  et~al., 2018, \mn@doi [\apjl] {10.3847/2041-8213/aab6ab}, \href
  {http://adsabs.harvard.edu/abs/2018ApJ...856L..23K} {856, L23}

\bibitem[\protect\citeauthoryear{{Kashikawa}, {Misawa}, {Minowa}, {Okoshi},
  {Hattori}, {Toshikawa}, {Ishikawa}  \& {Onoue}}{{Kashikawa}
  et~al.}{2014}]{Kashikawa2014}
{Kashikawa} N.,  {Misawa} T.,  {Minowa} Y.,  {Okoshi} K.,  {Hattori} T.,
  {Toshikawa} J.,  {Ishikawa} S.,   {Onoue} M.,  2014, \mn@doi [\apj]
  {10.1088/0004-637X/780/2/116}, \href
  {http://adsabs.harvard.edu/abs/2014ApJ...780..116K} {780, 116}

\bibitem[\protect\citeauthoryear{{Kennicutt}}{{Kennicutt}}{1998}]{Kennicutt1998}
{Kennicutt} Robert~C. J.,  1998, \mn@doi [\araa]
  {10.1146/annurev.astro.36.1.189}, \href
  {https://ui.adsabs.harvard.edu/abs/1998ARA&A..36..189K} {36, 189}

\bibitem[\protect\citeauthoryear{{Kochiashvili} et~al.,}{{Kochiashvili}
  et~al.}{2015}]{Kochiashvili2015}
{Kochiashvili} I.,  et~al., 2015, \mn@doi [\aap] {10.1051/0004-6361/201425535},
  \href {https://ui.adsabs.harvard.edu/abs/2015A&A...580A..42K} {580, A42}

\bibitem[\protect\citeauthoryear{{Komatsu} et~al.,}{{Komatsu}
  et~al.}{2011}]{Komatsu2011}
{Komatsu} E.,  et~al., 2011, \mn@doi [\apjs] {10.1088/0067-0049/192/2/18},
  \href {http://adsabs.harvard.edu/abs/2011ApJS..192...18K} {192, 18}

\bibitem[\protect\citeauthoryear{{Krogager}}{{Krogager}}{2018}]{Krogager2018}
{Krogager} J.-K.,  2018, arXiv e-prints, \href
  {https://ui.adsabs.harvard.edu/abs/2018arXiv180301187K} {p. arXiv:1803.01187}

\bibitem[\protect\citeauthoryear{{Krogager}, {Fynbo}, {M{\o}ller}, {Ledoux},
  {Noterdaeme}, {Christensen}, {Milvang-Jensen}  \& {Sparre}}{{Krogager}
  et~al.}{2012}]{Krogager2012}
{Krogager} J.-K.,  {Fynbo} J.~P.~U.,  {M{\o}ller} P.,  {Ledoux} C.,
  {Noterdaeme} P.,  {Christensen} L.,  {Milvang-Jensen} B.,   {Sparre} M.,
  2012, \mn@doi [\mnras] {10.1111/j.1745-3933.2012.01272.x}, \href
  {http://adsabs.harvard.edu/abs/2012MNRAS.424L...1K} {424, L1}

\bibitem[\protect\citeauthoryear{{Krogager} et~al.,}{{Krogager}
  et~al.}{2013}]{Krogager2013}
{Krogager} J.-K.,  et~al., 2013, \mn@doi [\mnras] {10.1093/mnras/stt955}, \href
  {http://adsabs.harvard.edu/abs/2013MNRAS.433.3091K} {433, 3091}

\bibitem[\protect\citeauthoryear{{Krogager}, {M{\o}ller}, {Fynbo}  \&
  {Noterdaeme}}{{Krogager} et~al.}{2017}]{Krogager2017}
{Krogager} J.-K.,  {M{\o}ller} P.,  {Fynbo} J.~P.~U.,   {Noterdaeme} P.,  2017,
  \mn@doi [\mnras] {10.1093/mnras/stx1011}, \href
  {http://adsabs.harvard.edu/abs/2017MNRAS.469.2959K} {469, 2959}

\bibitem[\protect\citeauthoryear{{Kulkarni}, {Hill}, {Schneider}, {Weymann},
  {Storrie-Lombardi}, {Rieke}, {Thompson}  \& {Jannuzi}}{{Kulkarni}
  et~al.}{2000}]{Kulkarni2000}
{Kulkarni} V.~P.,  {Hill} J.~M.,  {Schneider} G.,  {Weymann} R.~J.,
  {Storrie-Lombardi} L.~J.,  {Rieke} M.~J.,  {Thompson} R.~I.,   {Jannuzi}
  B.~T.,  2000, \mn@doi [\apj] {10.1086/308904}, \href
  {http://adsabs.harvard.edu/abs/2000ApJ...536...36K} {536, 36}

\bibitem[\protect\citeauthoryear{{Laursen}, {Razoumov}  \&
  {Sommer-Larsen}}{{Laursen} et~al.}{2009}]{Laursen2009}
{Laursen} P.,  {Razoumov} A.~O.,   {Sommer-Larsen} J.,  2009, \mn@doi [\apj]
  {10.1088/0004-637X/696/1/853}, \href
  {http://adsabs.harvard.edu/abs/2009ApJ...696..853L} {696, 853}

\bibitem[\protect\citeauthoryear{{Livermore} et~al.,}{{Livermore}
  et~al.}{2015}]{Livermore2015}
{Livermore} R.~C.,  et~al., 2015, \mn@doi [\mnras] {10.1093/mnras/stv686},
  \href {https://ui.adsabs.harvard.edu/abs/2015MNRAS.450.1812L} {450, 1812}

\bibitem[\protect\citeauthoryear{{Lopez}, {Reimers}, {D'Odorico}  \&
  {Prochaska}}{{Lopez} et~al.}{2002}]{Lopez2002}
{Lopez} S.,  {Reimers} D.,  {D'Odorico} S.,   {Prochaska} J.~X.,  2002, \mn@doi
  [\aap] {10.1051/0004-6361:20020181}, \href
  {http://adsabs.harvard.edu/abs/2002A%26A...385..778L} {385, 778}

\bibitem[\protect\citeauthoryear{{L{\'o}pez} et~al.,}{{L{\'o}pez}
  et~al.}{2016}]{Lopez2016}
{L{\'o}pez} S.,  et~al., 2016, \mn@doi [\aap] {10.1051/0004-6361/201628161},
  \href {https://ui.adsabs.harvard.edu/abs/2016A%26A...594A..91L} {594, A91}

\bibitem[\protect\citeauthoryear{{Modigliani} et~al.,}{{Modigliani}
  et~al.}{2010}]{modigliani2010}
{Modigliani} A.,  et~al., 2010, in Observatory Operations: Strategies,
  Processes, and Systems III. p. 773728, \mn@doi{10.1117/12.857211}

\bibitem[\protect\citeauthoryear{{M{\o}ller} \& {Christensen}}{{M{\o}ller} \&
  {Christensen}}{2020}]{Moller2020}
{M{\o}ller} P.,  {Christensen} L.,  2020, \mn@doi [\mnras]
  {10.1093/mnras/staa128}, \href
  {https://ui.adsabs.harvard.edu/abs/2020MNRAS.492.4805M} {492, 4805}

\bibitem[\protect\citeauthoryear{{M{\o}ller}, {Warren}, {Fall}, {Fynbo}  \&
  {Jakobsen}}{{M{\o}ller} et~al.}{2002}]{Moller2002}
{M{\o}ller} P.,  {Warren} S.~J.,  {Fall} S.~M.,  {Fynbo} J.~U.,   {Jakobsen}
  P.,  2002, \mn@doi [\apj] {10.1086/340934}, \href
  {https://ui.adsabs.harvard.edu/abs/2002ApJ...574...51M} {574, 51}

\bibitem[\protect\citeauthoryear{{M{\o}ller}, {Fynbo}, {Ledoux}  \&
  {Nilsson}}{{M{\o}ller} et~al.}{2013}]{Moller2013}
{M{\o}ller} P.,  {Fynbo} J.~P.~U.,  {Ledoux} C.,   {Nilsson} K.~K.,  2013,
  \mn@doi [\mnras] {10.1093/mnras/stt067}, \href
  {http://adsabs.harvard.edu/abs/2013MNRAS.430.2680M} {430, 2680}

\bibitem[\protect\citeauthoryear{{M{\o}ller} et~al.,}{{M{\o}ller}
  et~al.}{2018}]{Moller2018}
{M{\o}ller} P.,  et~al., 2018, \mn@doi [\mnras] {10.1093/mnras/stx2845}, \href
  {http://adsabs.harvard.edu/abs/2018MNRAS.474.4039M} {474, 4039}

\bibitem[\protect\citeauthoryear{{Noterdaeme} et~al.,}{{Noterdaeme}
  et~al.}{2012}]{Noterdaeme2012}
{Noterdaeme} P.,  et~al., 2012, \mn@doi [\aap] {10.1051/0004-6361/201220259},
  \href {http://adsabs.harvard.edu/abs/2012A%26A...547L...1N} {547, L1}

\bibitem[\protect\citeauthoryear{{Peng}, {Ho}, {Impey}  \& {Rix}}{{Peng}
  et~al.}{2002}]{Peng2002}
{Peng} C.~Y.,  {Ho} L.~C.,  {Impey} C.~D.,   {Rix} H.-W.,  2002, \mn@doi [\aj]
  {10.1086/340952}, \href {http://adsabs.harvard.edu/abs/2002AJ....124..266P}
  {124, 266}

\bibitem[\protect\citeauthoryear{{P{\'e}roux}, {Dessauges-Zavadsky},
  {D'Odorico}, {Kim}  \& {McMahon}}{{P{\'e}roux} et~al.}{2003}]{Peroux2003}
{P{\'e}roux} C.,  {Dessauges-Zavadsky} M.,  {D'Odorico} S.,  {Kim} T.-S.,
  {McMahon} R.~G.,  2003, \mn@doi [\mnras] {10.1046/j.1365-8711.2003.06952.x},
  \href {http://adsabs.harvard.edu/abs/2003MNRAS.345..480P} {345, 480}

\bibitem[\protect\citeauthoryear{{P{\'e}roux}, {Bouch{\'e}}, {Kulkarni}, {York}
   \& {Vladilo}}{{P{\'e}roux} et~al.}{2012}]{Peroux2012}
{P{\'e}roux} C.,  {Bouch{\'e}} N.,  {Kulkarni} V.~P.,  {York} D.~G.,
  {Vladilo} G.,  2012, \mn@doi [\mnras] {10.1111/j.1365-2966.2011.19947.x},
  \href {http://adsabs.harvard.edu/abs/2012MNRAS.419.3060P} {419, 3060}

\bibitem[\protect\citeauthoryear{{Prochaska}, {Gawiser}, {Wolfe}, {Castro}  \&
  {Djorgovski}}{{Prochaska} et~al.}{2003}]{Prochaska2003}
{Prochaska} J.~X.,  {Gawiser} E.,  {Wolfe} A.~M.,  {Castro} S.,   {Djorgovski}
  S.~G.,  2003, \mn@doi [\apjl] {10.1086/378945}, \href
  {http://adsabs.harvard.edu/abs/2003ApJ...595L...9P} {595, L9}

\bibitem[\protect\citeauthoryear{{Prochaska}, {Herbert-Fort}  \&
  {Wolfe}}{{Prochaska} et~al.}{2005}]{Prochaska2005}
{Prochaska} J.~X.,  {Herbert-Fort} S.,   {Wolfe} A.~M.,  2005, \mn@doi [\apj]
  {10.1086/497287}, \href {http://adsabs.harvard.edu/abs/2005ApJ...635..123P}
  {635, 123}

\bibitem[\protect\citeauthoryear{{Rafelski}, {Neeleman}, {Fumagalli}, {Wolfe}
  \& {Prochaska}}{{Rafelski} et~al.}{2014}]{Rafelski2014}
{Rafelski} M.,  {Neeleman} M.,  {Fumagalli} M.,  {Wolfe} A.~M.,   {Prochaska}
  J.~X.,  2014, \mn@doi [\apjl] {10.1088/2041-8205/782/2/L29}, \href
  {http://adsabs.harvard.edu/abs/2014ApJ...782L..29R} {782, L29}

\bibitem[\protect\citeauthoryear{{Rhodin}, {Christensen}, {M{\o}ller}, {Zafar}
  \& {Fynbo}}{{Rhodin} et~al.}{2018}]{Rhodin2018}
{Rhodin} N.~H.~P.,  {Christensen} L.,  {M{\o}ller} P.,  {Zafar} T.,   {Fynbo}
  J.~P.~U.,  2018, \mn@doi [\aap] {10.1051/0004-6361/201832992}, \href
  {https://ui.adsabs.harvard.edu/abs/2018A%26A...618A.129R} {618, A129}

\bibitem[\protect\citeauthoryear{{Rhodin}, {Agertz}, {Christensen}, {Renaud}
  \& {Fynbo}}{{Rhodin} et~al.}{2019}]{Rhodin2019}
{Rhodin} N.~H.~P.,  {Agertz} O.,  {Christensen} L.,  {Renaud} F.,   {Fynbo}
  J.~P.~U.,  2019, \mn@doi [\mnras] {10.1093/mnras/stz1479}, \href
  {https://ui.adsabs.harvard.edu/abs/2019MNRAS.488.3634R} {488, 3634}

\bibitem[\protect\citeauthoryear{{STSCI Development Team}}{{STSCI Development
  Team}}{2012}]{Drizzlepac2012}
{STSCI Development Team} 2012, {DrizzlePac: HST image software}, Astrophysics
  Source Code Library (\mn@eprint {ascl} {1212.011})

\bibitem[\protect\citeauthoryear{{S{\'a}nchez-Ram{\'{\i}}rez}
  et~al.,}{{S{\'a}nchez-Ram{\'{\i}}rez} et~al.}{2016}]{SanchezRamirez2016}
{S{\'a}nchez-Ram{\'{\i}}rez} R.,  et~al., 2016, \mn@doi [\mnras]
  {10.1093/mnras/stv2732}, \href
  {http://adsabs.harvard.edu/abs/2016MNRAS.456.4488S} {456, 4488}

\bibitem[\protect\citeauthoryear{{Schlafly} \& {Finkbeiner}}{{Schlafly} \&
  {Finkbeiner}}{2011}]{Schlafly2011}
{Schlafly} E.~F.,  {Finkbeiner} D.~P.,  2011, \mn@doi [\apj]
  {10.1088/0004-637X/737/2/103}, \href
  {http://adsabs.harvard.edu/abs/2011ApJ...737..103S} {737, 103}

\bibitem[\protect\citeauthoryear{{Verhamme}, {Schaerer}, {Atek}  \&
  {Tapken}}{{Verhamme} et~al.}{2008}]{Verhamme2008}
{Verhamme} A.,  {Schaerer} D.,  {Atek} H.,   {Tapken} C.,  2008, \mn@doi [\aap]
  {10.1051/0004-6361:200809648}, \href
  {http://adsabs.harvard.edu/abs/2008A%26A...491...89V} {491, 89}

\bibitem[\protect\citeauthoryear{{Vernet} et~al.,}{{Vernet}
  et~al.}{2011}]{Vernet2011}
{Vernet} J.,  et~al., 2011, \mn@doi [\aap] {10.1051/0004-6361/201117752}, \href
  {http://adsabs.harvard.edu/abs/2011A%26A...536A.105V} {536, A105}

\bibitem[\protect\citeauthoryear{{Warren}, {M{\o}ller}, {Fall}  \&
  {Jakobsen}}{{Warren} et~al.}{2001}]{Warren2001}
{Warren} S.~J.,  {M{\o}ller} P.,  {Fall} S.~M.,   {Jakobsen} P.,  2001, \mn@doi
  [\mnras] {10.1046/j.1365-8711.2001.04629.x}, \href
  {http://adsabs.harvard.edu/abs/2001MNRAS.326..759W} {326, 759}

\bibitem[\protect\citeauthoryear{{Weatherley}, {Warren}, {M{\o}ller}, {Fall},
  {Fynbo}  \& {Croom}}{{Weatherley} et~al.}{2005}]{Weatherley2005}
{Weatherley} S.~J.,  {Warren} S.~J.,  {M{\o}ller} P.,  {Fall} S.~M.,  {Fynbo}
  J.~U.,   {Croom} S.~M.,  2005, \mn@doi [\mnras]
  {10.1111/j.1365-2966.2005.08838.x}, \href
  {http://adsabs.harvard.edu/abs/2005MNRAS.358..985W} {358, 985}

\bibitem[\protect\citeauthoryear{{Wenger} et~al.,}{{Wenger}
  et~al.}{2000}]{Simbad2000}
{Wenger} M.,  et~al., 2000, \mn@doi [\aaps] {10.1051/aas:2000332}, \href
  {http://adsabs.harvard.edu/abs/2000A%26AS..143....9W} {143, 9}

\bibitem[\protect\citeauthoryear{{Whitaker} et~al.,}{{Whitaker}
  et~al.}{2014}]{Whitaker2014}
{Whitaker} K.~E.,  et~al., 2014, \mn@doi [\apj] {10.1088/0004-637X/795/2/104},
  \href {http://adsabs.harvard.edu/abs/2014ApJ...795..104W} {795, 104}

\bibitem[\protect\citeauthoryear{{Whitaker} et~al.,}{{Whitaker}
  et~al.}{2015}]{Whitaker2015}
{Whitaker} K.~E.,  et~al., 2015, \mn@doi [\apjl] {10.1088/2041-8205/811/1/L12},
  \href {http://adsabs.harvard.edu/abs/2015ApJ...811L..12W} {811, L12}

\bibitem[\protect\citeauthoryear{{Wolfe}, {Turnshek}, {Smith}  \&
  {Cohen}}{{Wolfe} et~al.}{1986}]{Wolfe1986}
{Wolfe} A.~M.,  {Turnshek} D.~A.,  {Smith} H.~E.,   {Cohen} R.~D.,  1986,
  \mn@doi [\apjs] {10.1086/191114}, \href
  {http://adsabs.harvard.edu/abs/1986ApJS...61..249W} {61, 249}

\bibitem[\protect\citeauthoryear{{Wolfe}, {Prochaska}, {Jorgenson}  \&
  {Rafelski}}{{Wolfe} et~al.}{2008}]{Wolfe2008}
{Wolfe} A.~M.,  {Prochaska} J.~X.,  {Jorgenson} R.~A.,   {Rafelski} M.,  2008,
  \mn@doi [\apj] {10.1086/588090}, \href
  {http://adsabs.harvard.edu/abs/2008ApJ...681..881W} {681, 881}

\bibitem[\protect\citeauthoryear{{Zafar~}, {Popping}  \& {P{\'e}roux}}{{Zafar~}
  et~al.}{2013}]{Zafar2013}
{Zafar~} T.,  {Popping} A.,   {P{\'e}roux} C.,  2013, \mn@doi [\aap]
  {10.1051/0004-6361/201321153}, \href
  {https://ui.adsabs.harvard.edu/abs/2013A&A...556A.140Z} {556, 140}

\bibitem[\protect\citeauthoryear{{Zafar}, {M{\o}ller}, {P{\'e}roux}, {Quiret},
  {Fynbo}, {Ledoux}  \& {Deharveng}}{{Zafar} et~al.}{2017}]{Zafar2017}
{Zafar} T.,  {M{\o}ller} P.,  {P{\'e}roux} C.,  {Quiret} S.,  {Fynbo} J.~P.~U.,
   {Ledoux} C.,   {Deharveng} J.-M.,  2017, \mn@doi [\mnras]
  {10.1093/mnras/stw2907}, \href
  {http://adsabs.harvard.edu/abs/2017MNRAS.465.1613Z} {465, 1613}

\bibitem[\protect\citeauthoryear{{van der Wel} et~al.,}{{van der Wel}
  et~al.}{2014}]{vanderWel2014}
{van der Wel} A.,  et~al., 2014, \mn@doi [\apj] {10.1088/0004-637X/788/1/28},
  \href {http://adsabs.harvard.edu/abs/2014ApJ...788...28V} {788, 28}

\makeatother
\end{thebibliography}



\appendix

\section{Absorption analysis of the DLA towards Q0139--0824}
\label{subsec:absanalysis}

The DLA towards Q0139--0824 has a reported absorption metallicity of
$[{\rm Si/H}]_{\rm abs} = -1.15 \pm 0.15$ \citep{Krogager2012,
  Hartoog2015}. On closer examination, this value appears to have
emerged based on unpublished measurements, whereas an extensive
literature search only returned a published iron abundance $[{\rm
    Fe/H}]_{\rm abs} = -1.62 \pm 0.02$, and a minimally depleted
metallicity of $[{\rm M/H}]_{\rm abs} = -1.27 \pm 0.19$
\citep{Wolfe2008}.  Despite the low ($\sim 0.5\sigma$) tension between
the $[{\rm Si/H}]$ and the $[{\rm M/H}]$ metallicity measurements, the
analysis of the former remains unpublished, and it is unclear which
element was used to determine the metallicity by
\citet{Wolfe2008}. These considerations motivate us to re-examine the
metallicity.

We use existing X-shooter \citep{Vernet2011} data (programme ID
088.A-0378; PI: Christensen) and perform an independent absorption
analysis to determine the gas-phase metallicity. The quasar was
observed with X-Shooter on 2012 January 7 with an integration time of
$2\times1800$~s on target. Slit widths of 1.3~arcsec and 1.2~arcsec
were used for the UVB and VIS+NIR arms, respectively. However, here we
only use the data from the UVB and VIS arms covering wavelengths from
$\sim$3000 to $\sim$10,000~\AA.  The data were reduced using the
official X-Shooter data processing pipeline version 2.9.3
\citep{modigliani2010} and flux calibrated using observations of the
spectrophotometric standard star, Feige 110, observed on the same
night.  The two exposures from the UVB and VIS arms were reduced
separately in `stare' mode, and the 1D spectra were optimally
extracted using custom routines and combined by weighting each
spectrum with the average signal to noise ratio.  To compute an
accurate value for the instrument resolution, we measured spectral
resolutions from telluric absorption lines in the quasar spectra to be
$\mathcal{R}_{\rm VIS} = 42$~km~s$^{-1}$.  Due to the lack of telluric
absorption features in the UVB spectrum, we scale its nominal
resolution to the ratio of observed-to-nominal resolution in the
visual arm, giving $\mathcal{R}_{\rm UVB} = 70$~km~s$^{-1}$,
consistent with $\mathcal{R} = 4100 = 73$~km~s$^{-1}$ reported for the
UVB arm of X-shooter given a slit width of 1.3~arcsec
\citep{Vernet2011}.

To infer the gas-phase metallicity of the $z=2.6773$ DLA, we measure
the column densities of low-ionisation metal absorption lines free of
telluric contamination. The absorption lines are fitted using a
Voigt-profile decomposition in order to alleviate the effects of
hidden saturation in the rather low-resolution data. We use the Python
package {\tt VoigtFit} \citep{Krogager2018}, which uses line-lists
with updated oscillator strengths by \citet{Cashman2017}. We fit all
available iron and silicon lines that are not blended with other
absorption features. For each line, we fit two components together
with the local continuum level around each line. Since the effective
broadening of lines from heavy elements such as Fe and Si are
dominated by turbulence rather than thermal motion, we tie the
redshifts and broadening parameters component-wise. The continuum
level is fitted simultaneously using a Chebyshev polynomium of second
order.

\begin{figure*}
\includegraphics[width=1.\textwidth]{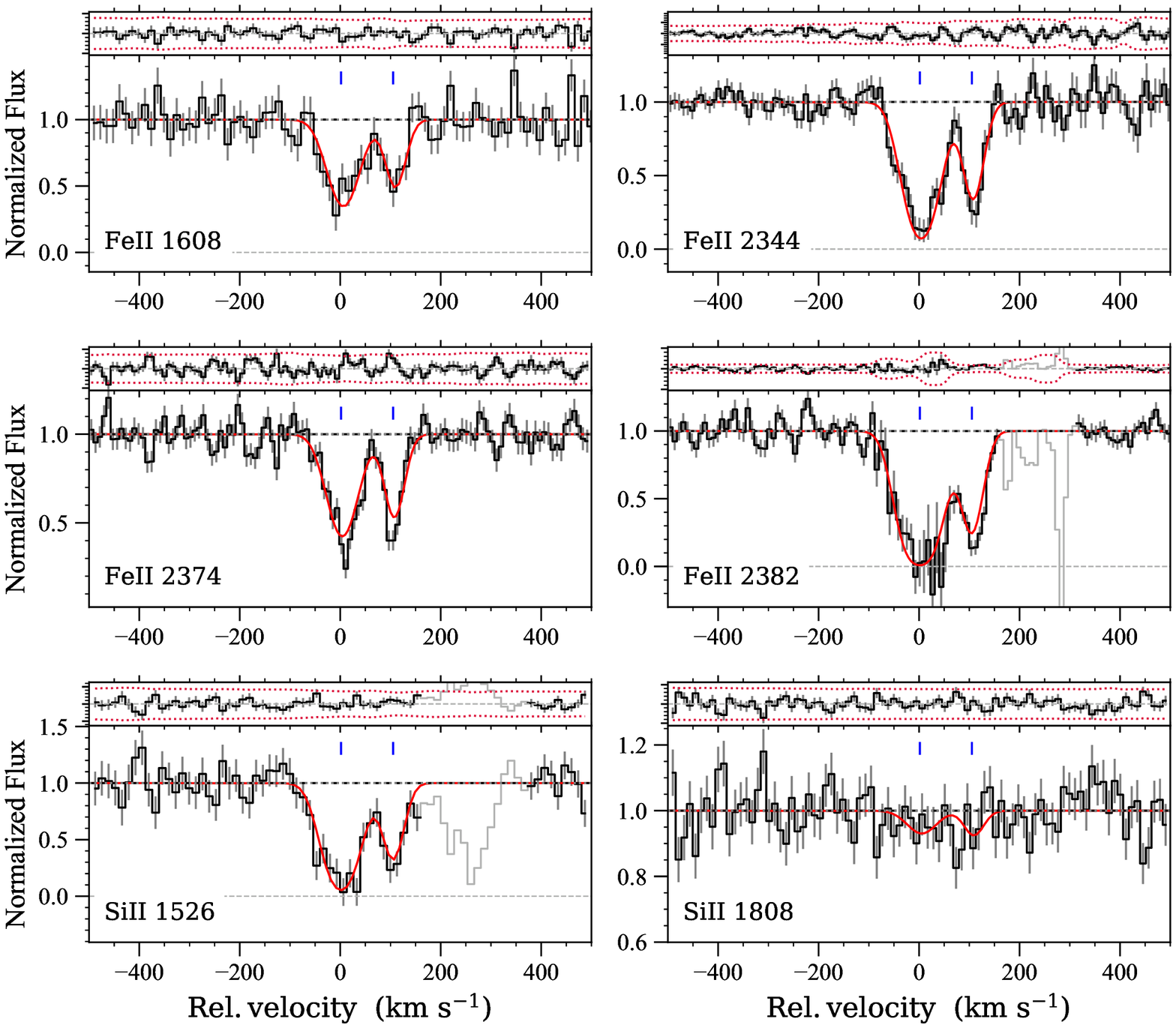}
\caption{Voigt profile fits to the low-ionisation metal absorption
  lines of the $z=2.6773$ DLA towards Q\,0139--0824. Each panel shows
  the continuum normalised flux of a single transition
  (\textit{bottom}), and the fit residuals (\textit{top}). The black
  line represents the observed spectrum with uncertainties shown as
  grey vertical bars. Grey regions without uncertainties were excluded
  in the fit. The red curve represents the best fit.}
  \label{fig:absanal}
\end{figure*}

The best-fit relative velocities, broadening parameters, and column
densities for each velocity component together with the total column
densities are reported in Table~\ref{tab:absanal} and the best-fit
line profiles are shown in Fig.~\ref{fig:absanal}.  We obtain a
measurement of the neutral hydrogen column density in perfect
agreement with the results by \citet{Wolfe2008} who find $\logNHIcm =
20.70\pm 0.15$. We infer estimates of the gas-phase metallicity of
$[{\rm Si/H}] = -1.2\pm 0.2$ and $[{\rm Fe/H}] = -1.4\pm 0.2$ using
solar photospheric abundances by \citet{Asplund2009}.

Because Fe is a highly refractory element and is very often strongly
depleted into dust grains, [Fe/H] does not serve as a robust indicator
of the gas-phase metallicity. Si is less depleted but may still
underestimate the overall metallicity. Less refractory elements such
as S and Zn are more suitable tracers of the gas-phase metallicity;
However, the three lines of \ion{S}{ii} at rest-frame wavelengths
1250, 1253 and 1259 \AA\ are strongly blended with Ly$\alpha$ forest
lines, and \ion{Zn}{ii}~$\lambda\lambda 2026,2062$\,\AA\ show a
different velocity profile most likely caused by contamination from
proximate telluric lines, or possibly also arising from a blend with
an intervening, unrelated absorption system. Yet, we only identify an
intervening \ion{C}{iv} absorber at $z_{\rm abs} = 2.36$, and a
possible \ion{Mg}{ii} absorber at $z_{\rm abs} = 2.23$ -- neither of
which produce strong lines at the position of the \ion{Zn}{ii} lines
from the $z=2.6773$ DLA.

We therefore report the metallicity of the absorbing gas based on
[Si/H]. This result is based on two transitions; the saturated
\ion{Si}{ii}~$ \lambda 1526$\,\AA\ line and the weak
\ion{Si}{ii}~$\lambda 1808$\,\AA\ line (see
Fig. \ref{fig:absanal}). However, since the broadening parameter is
tied to that of \ion{Fe}{ii}, we still probe different lines with
oscillator strengths spanning an order of magnitude. Therefore, the
relative strength of all the fitted lines effectively constrain both
the column density and the broadening parameter. The total column
densities should therefore be robust, however, higher spectral
resolution data are needed to firmly rule out any significant hidden
saturation.  In conclusion, the results of our absorption analysis are
consistent with the literature values to within 1$\sigma$ for both the
iron and silicon based measurements. We report a silicon-based
metallicity of $[{\rm Si/H}] = -1.2 \pm 0.2$, which may be slightly
higher due to dust depletion and hidden saturation. The absorber
therefore still meets the metallicity cut in our sample selection when
considering the statistical and systematic uncertainties.

\begin{table}
\caption{Absorption analysis of the low-ionisation metal lines
of the DLA towards Q0139--0824 at $z_{\rm abs} = 2.6773$.}
\label{tab:absanal}
\begin{tabular}{llccccc}
\hline
Complex & $b$ & $\log _{10}({\rm N~/~cm}^{-2})$ \\
$\lambda$  = [{\AA}] ,  $v$ = [km~s$^{-1}$] & [km~s$^{-1}$] & & \\
\hline
Fe\,\textsc{ii}~$\lambda\lambda 1608,2344,2374,2382$ \\
$v_1 = 0\pm 1$ & $32\pm2$ & $14.51\pm0.03$ \\
$v_2 = 104\pm 2$ & $11\pm1$ & $14.39\pm0.07$ \\
Total Column: & & $14.76\pm0.04$\\

Si\,\textsc{ii}~$\lambda\lambda 1526,1808$ \\
$v_1 = 0\pm 1$ & $32\pm2$ & $14.71\pm0.14$ \\
$v_2 = 104\pm 2$ & $11\pm1$ & $14.59\pm0.20$ \\
Total Column: & & $14.97\pm0.12$\\
\hline
\end{tabular}
\end{table}

\bsp	
\label{lastpage}
\end{document}